\documentclass[%
 reprint,
superscriptaddress,
 amsmath,amssymb,
 aps,
 prl, 
floatfix,
]{revtex4-2}

\usepackage{graphicx}
\usepackage{dcolumn}
\usepackage{bm}
\usepackage{svg}
\usepackage[normalem]{ulem}
\usepackage{hyperref}

\begin{document}

\preprint{APS/123-QED}

\title{Engineering multi-photon dissipation with a dc-voltage-biased Josephson junction}

\author{Marco Paradina}
\affiliation{\'Ecole Normale Sup\'erieure de Lyon,  CNRS, Laboratoire de Physique, F-69342 Lyon, France}
\affiliation{Alice \& Bob, 53 Bd du G\'en\'eral Martial Valin, 75015 Paris, France}

\author{Ambroise Peugeot}%
\email[Contact author: ]{ambroise.peugeot@ens-lyon.fr}
\affiliation{\'Ecole Normale Sup\'erieure de Lyon,  CNRS, Laboratoire de Physique, F-69342 Lyon, France}

\author{Roberto Negrin}
\affiliation{Alice \& Bob, 53 Bd du G\'en\'eral Martial Valin, 75015 Paris, France}
\affiliation{Laboratoire de Physique de l’\'Ecole Normale Supérieure, Mines Paris, Inria, CNRS, ENS-PSL, Centre Automatique et Systèmes (CAS),
Sorbonne Université, PSL Research University, Paris, France}

\author{Oscar Novat}%
\affiliation{\'Ecole Normale Sup\'erieure de Lyon,  CNRS, Laboratoire de Physique, F-69342 Lyon, France}

\author{Tristan Villain}%
\affiliation{\'Ecole Normale Sup\'erieure de Lyon,  CNRS, Laboratoire de Physique, F-69342 Lyon, France}

\author{Anil Murani}
\affiliation{Alice \& Bob, 53 Bd du G\'en\'eral Martial Valin, 75015 Paris, France}

\author{Jean-Loup Ville}
\affiliation{Alice \& Bob, 53 Bd du G\'en\'eral Martial Valin, 75015 Paris, France}

\author{Sébastien Jezouin}
\affiliation{Alice \& Bob, 53 Bd du G\'en\'eral Martial Valin, 75015 Paris, France}

\author{Raphaël Lescanne}
\affiliation{Alice \& Bob, 53 Bd du G\'en\'eral Martial Valin, 75015 Paris, France}

\author{Audrey Bienfait}%
\affiliation{\'Ecole Normale Sup\'erieure de Lyon,  CNRS, Laboratoire de Physique, F-69342 Lyon, France}

\author{Benjamin Huard}%
\affiliation{\'Ecole Normale Sup\'erieure de Lyon,  CNRS, Laboratoire de Physique, F-69342 Lyon, France}
 
\date{\today}

\begin{abstract}
Multi-photon dissipation -- a key resource for bosonic qubits -- is usually realized by parametrically pumping a Josephson coupler at the cost of spurious nonlinear terms. Here we instead engineer it using a dc-voltage-biased SQUID, such that these parasitic terms average out. We activate the conversion of one, two, or four photons of a high-Q mode into a single photon of a lossy mode. We characterize the two-photon dissipation by Wigner tomography, establishing dc-biased junctions as a resource for reservoir engineering and a viable route to cat-qubit stabilization.
\end{abstract}

\maketitle

The unavoidable coupling of a quantum system to its environment leads to its decoherence. Counterintuitively, dissipation can also be turned into a resource: quantum reservoir engineering harnesses dissipation to steer open quantum systems toward a desired state or state manifold~\cite{Poyatos1996,harrington_engineered_2022}. The standard ingredient is a lossy mode --- the buffer --- at frequency $\omega_b$, which rapidly cools down owing to its coupling to the environment. By coupling the system of interest to this buffer in a controllable way, one engineers an effective dissipation channel that evacuates entropy. In the context of harmonic oscillators, reservoir engineering can tune their decay rate~\cite{schliesser_resolved-sideband_2008,rocheleau_preparation_2010,teufel_sideband_2011,Pfaff2017,blumenthal_experimental_2026,barge_temporal_2026}, stabilize vacuum squeezed states~\cite{cirac_dark_1993,wollman_quantum_2015,pirkkalainen_squeezing_2015,lecocq_quantum_2015,lei_quantum_2016,kienzler_quantum_2015,dassonneville_dissipative_2021} or entanglement~\cite{krauter_entanglement_2011,woolley_two-mode_2014}, or stabilize number states~\cite{holland_single-photon-resolved_2015,Souquet2016,li_autonomous_2024}. It can be used for quantum error correction by stabilizing qubit subspaces with two-photon dissipation~\cite{wolinsky_quantum_1988,Mirrahimi2014,leghtas_confining_2015,Touzard2017,lescanne2020exponential,marquet_harnessing_2024} or more elaborate dissipative schemes~\cite{Gertler2020a,lachance-quirion_autonomous_2024,sellem_dissipative_2025}.

More specifically, in superconducting circuits, multi-photon dissipation can be achieved using Josephson-junction-based couplers between a memory mode $a$ of interest and the buffer mode $b$. Several parametrically pumped couplers~\cite{Bergeal2010b,Frattini2017,lescanne2020exponential,Ye2020,lu_high-fidelity_2023,maiti_linear_2025,bhandari_symmetrically_2025,hua_engineering_2025,vanselow_dissipating_2026} and an autoparametric one~\cite{Marquet2024,marquet_harnessing_2024} have been designed to isolate the desired coupling term $\hbar g_n(\hat{a}^n\hat{b}^\dagger +(\hat{a}^\dagger)^n\hat{b})$ and minimize the contribution of parasitic non-linear terms arising from the Josephson Hamiltonian~\cite{Nigg2012a}. However, in all these designs, stray inductances and the fabrication uncertainties of Josephson junctions prevent the complete suppression of the self-Kerr effect on mode $a$, the cross-Kerr effect between modes $a$ and $b$, and a variety of other detrimental effects for quantum control, particularly at large pump power~\cite{putterman_preserving_2025,carde_flux-pump_2026}. Overcoming this fundamental limitation, rather than mitigating it, calls for a coupling mechanism in which these parasitic terms vanish by construction.

\begin{figure}[tb]
  \includegraphics[width=\columnwidth]{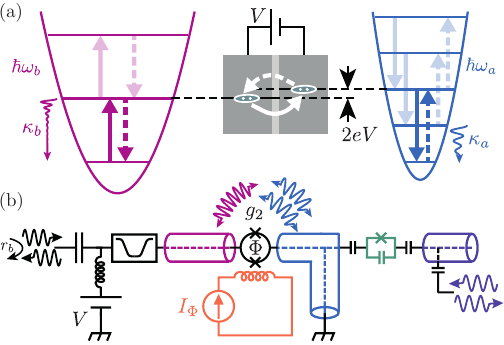}
  \caption{\label{fig:fig1}(a) Energy levels of memory mode $a$ (blue) and buffer mode $b$ (magenta) coupled to a Josephson junction biased with a voltage $V$. For $2eV=2\hbar\omega_a-\hbar\omega_b$, Cooper pair tunneling enables transitions converting a pair of photons of mode $a$ into a single photon of mode $b$ (solid arrows) and back (dashed arrows).
(b) Circuit schematic. Memory and buffer are $\lambda/4$ resonators coupled to a SQUID, acting as a single junction with a flux tunable Josephson energy $E_J(\Phi)$. A stepped impedance filter inhibits memory decay through the buffer line while ensuring a galvanic connection as well as a large coupling rate for the buffer mode. A bias tee allows us to dc-voltage-bias the SQUID and perform RF measurements. At $2eV=2\hbar\omega_a-\hbar\omega_b$, memory photon pairs are exchanged with single buffer photons at a rate $g_2$. A transmon qubit (green) is dispersively coupled to both a $\lambda/2$ readout resonator (purple) and the memory for quantum state tomography.}
\end{figure}

In parallel, dc-voltage-biased Josephson junctions have emerged as a way to manipulate microwave modes~\cite{Hofheinz2011,Westig2017,Jebari2018,Grimm2019,Rolland2019,Peugeot2021,Menard2022,aiello_quantum_2022,albert_microwave_2024}. In contrast to the usual multimode mixing induced by parametric pumping, the pump photon energy is replaced by the energy $2eV$ gained or lost by Cooper pairs as they tunnel across the junction biased with a voltage $V$ (Fig.~\ref{fig:fig1}a). In the context of a photomultiplier~\cite{leppakangas2018multiplying}, it was shown~\cite{albert_microwave_2024,privatecomm} that the 2-to-1 photon conversion rate can be as large as $g_2/2\pi\approx 70~\mathrm{MHz}$. A different mechanism based on quasiparticle tunneling was shown to activate dissipation of packets of $n$ photons at rates beyond $100~\mathrm{MHz}$ in a lossy resonator~\cite{aiello_quantum_2022}, at the cost of activating the process for all values of $n$ simultaneously. Recent proposals~\cite{aissaoui_cat-qubit-stabilization_2026,rojkov2026stabilization} have identified the Cooper pair tunneling approach as a way to generate multi-photon dissipation and eventually stabilize cat qubits without the parasitic nonlinear terms that limit pumped couplers. This dc-bias route offers further practical advantages: the absence of a pump suppresses harmonics that could excite spurious higher modes, and a dc-voltage bias is more easily scaled than a microwave drive.

In this Letter, we demonstrate how a dc-voltage-biased SQUID can engineer single- and two-photon loss for a high-Q memory resonator (Fig.~\ref{fig:fig1}a), as well as activate up to 4-to-1 photon swaps with a buffer mode. The circuit is designed to apply a dc-voltage bias on the SQUID without limiting the lifetime of the memory mode $a$ to which it is connected. A caveat is that voltage noise could smear out the matching between $2eV$ and the energy difference between $n$ memory photons and one buffer photon. We show that proper filtering reduces it to values that do not affect multi-photon dissipation engineering. Using a transmon qubit to probe the quantum state of mode $a$ as a function of time, we characterize the engineered dissipation rate and various parasitic effects. Our experiment indicates that, with modest improvements in radiofrequency filtering, dc-voltage-biased Josephson junctions are a viable route to engineer pure multi-photon dissipation and ultimately stabilize cat qubits.

The circuit consists of buffer and memory coplanar-waveguide resonators galvanically connected to a SQUID (Fig.~\ref{fig:fig1}b). Up to a small junction asymmetry~\cite{supmat}, the SQUID behaves as a single Josephson junction whose energy $E_J(\Phi)={E_J}^\mathrm{max}|\cos(\pi\Phi/\Phi_0)|$ is tuned in situ through a fast-flux line, with $\Phi_0=h/2e$ the superconducting flux quantum and a maximum ${E_J}^\mathrm{max}/h=13~\mathrm{GHz}$. A bias tee connects the buffer port both to the microwave measurement lines and to a low-noise bias line~\cite{supmat}, allowing us to dc-voltage-bias the junction. While maintaining this galvanic dc connection to the SQUID, an on-chip four-stage stepped-impedance band-stop filter sets the output coupling of the buffer mode $b$ ($\omega_b/2\pi=7.56~\mathrm{GHz}$) to a large rate $\kappa_b/2\pi= ~85 \pm 5~\mathrm{MHz}$, ensuring fast elimination of the buffer photons. The band-stop is centered around the memory frequency $\omega_a/2\pi=4.0429~\mathrm{GHz}$ with a bandwidth of 2~GHz so that it both limits the leakage rate of memory photons into the line to $2\pi\times 1~\mathrm{kHz}$ and suppresses spurious swaps of memory photons with modes of the bias lines~\cite{supmat}. Finally, the memory is coupled to a transmon qubit with a dispersive shift $\chi/2\pi=1.51~\mathrm{MHz}$, used to perform quantum state tomography. The circuit is patterned in a sputtered tantalum film on a sapphire substrate with aluminum Josephson junctions, following established recipes for long-lived resonators, and cooled to an 8~mK base temperature in a dilution refrigerator~\cite{supmat}.

The controlled dissipation relies on dynamical Coulomb blockade~\cite{devoret_effect_1990,girvin_quantum_1990,averin_incoherent_1990,ingold_charge_1992,Hofheinz2011}. The tunneling of a Cooper pair of charge $2e$ across the junction biased at voltage $V$ is necessarily accompanied by the release or absorption of an energy $2eV$ by the rest of the circuit (Fig.~\ref{fig:fig1}a). Setting $2eV=n\hbar\omega_a-\hbar\omega_b$ matches this energy to the conversion of $n$ memory photons into a single buffer photon. Following Ref.~\cite{leppakangas2018multiplying}, this process derives from the Josephson Hamiltonian
\begin{equation}
    H_{J} = -E_J \cos\left(  \phi_a (\hat{a}+\hat{a}^\dagger) + \phi_b (\hat{b}+\hat{b}^\dagger)-\omega_J t - \xi(t)\right),
\end{equation}
where $\hat a,\hat b$ are annihilation operators, $\phi_i = \sqrt{2e^2Z_i/\hbar}$ is the zero-point phase fluctuation of mode $i$ set by its characteristic impedance $Z_i$, here designed to be $\phi_a=0.18$ and $\phi_b=0.21$, $\omega_J = 2e\bar V/\hbar$ is the Josephson frequency imposed by the average bias $\bar V$, and the offset phase $\xi(t)=\int_{-\infty}^t 2e(V-\bar V)/\hbar\mathrm{d}t'$ grows commensurably with the bias fluctuations. Expanding the cosine to order $n+2$ and moving to the frame rotating at $\omega_J = n\omega_a - \omega_b$, a rotating-wave approximation gives
\begin{equation}
\label{Hamiltonian}
    H_{n} = -\hbar g_n e^{i\xi(t)} \hat{a}^n \hat{b}^\dagger + h.c.,
\end{equation}
up to corrections of order $\phi_i^{n+3}$, with $g_n = \frac{E_J}{2\hbar} \frac{\phi_a^n\phi_b}{n!}$ the exchange rate between $n$ memory photons and one buffer photon.

\begin{figure}[tb]
  \includegraphics[width=\columnwidth]{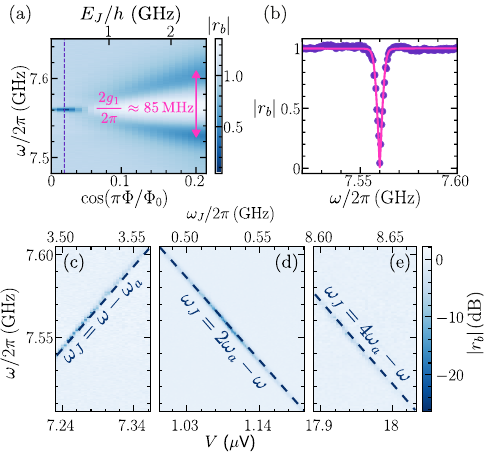}
  \caption{\label{fig:fig2}
  (a) Magnitude of the buffer reflection coefficient $|r_b|$ versus probe frequency $\omega$ and $\cos(\pi\Phi/\Phi_0)$ (equivalent $E_J(\Phi)$, top axis). The 1-to-1 photon swap is activated by setting $2eV = \hbar (\omega_b - \omega_a)= h \times 3.52 ~\mathrm{GHz} $. As the swap rate $g_1\propto|\cos(\pi\Phi/\Phi_0)|$ increases, the single absorption dip at $\omega_b$ splits owing to the hybridization of buffer and memory modes. The separation $2g_1$ reaches $2g_1/2\pi \approx 85~\mathrm{MHz}$.
  (b) Dots: reflection coefficient at critical coupling ($\cos(\pi\Phi/\Phi_0)=0.024$ as indicated in (a)). Solid line: fit using an input--output model that includes the residual bias-voltage noise~\cite{supmat}, giving $g_1/2\pi = 4.8~\mathrm{MHz}$ and a Josephson-frequency linewidth $\sigma_J/2\pi =  1.5~\mathrm{MHz}$.
  (c,d,e) $|r_b|$ versus bias voltage $V$ (equivalently $\omega_J$, top axis), set to activate a 1-to-1 (c, $E_J/h = 0.2~\mathrm{GHz}$), 2-to-1 (d, $E_J/h = 0.4~\mathrm{GHz}$) and 4-to-1 (e, $E_J/h = 0.5~\mathrm{GHz}$) photon swap between memory and buffer, evidenced by each absorption feature following its energy-matching condition (dashed lines).}
\end{figure}

Bias fluctuations $\delta V(t)= V(t) - \bar{V}$ make $\xi(t)$ diffuse randomly, imprinting an unknown phase on the converted photons through the $e^{i\xi(t)}\hat a^n\hat b^\dagger$ term, so this swap alone cannot coherently transfer a quantum state between $a$ and $b$ modes~\cite{aissaoui_cat-qubit-stabilization_2026,danner_quantum_2025}. It can however engineer dissipation. For a large enough buffer loss rate such that $\sigma_J=\frac{2e}{\hbar}\sqrt{\langle\delta V^2\rangle}\ll\kappa_b$~\cite{aissaoui_cat-qubit-stabilization_2026}, and $\kappa_b\gg 4 n g_n \langle \hat{a}^\dagger\hat{a}\rangle^{(n-1)/2}$, the mode $b$ can be adiabatically eliminated~\cite{lescanne2020exponential}, which yields a (multi)photon dissipator $\hat{L}_n=\sqrt{\kappa_n}\hat{a}^n$ on the memory, with
\begin{equation}
    \kappa_n = \frac{4g_n^2}{\kappa_b}.
\end{equation}
Moreover, the same $\cos(\omega_J t)$ modulation that activates this dissipator also dresses every spurious term --- the self-Kerr of mode $a$, the cross-Kerr between $a$ and $b$, and the bias-induced detunings --- which all rotate at $\omega_J$ and average to zero at first order in the rotating wave approximation, in contrast to pumped couplers where they remain static~\cite{aissaoui_cat-qubit-stabilization_2026}.

Microwave reflectometry first lets us identify the two modes. At zero bias, the SQUID acts as a flux-tunable inductive load at the voltage antinode of the resonators, so the mode frequencies depend on flux; we use this dependence to calibrate $\Phi$ as a function of the flux-line current $I_\Phi$ as well as ${E_J}^\mathrm{max}$~\cite{supmat}. At finite bias, however, this inductive detuning is one of the terms modulated at $\omega_J$ and averages to zero in the rotating frame, so the memory and buffer frequencies become flux-independent, pinned at the bare values given above. This flux-independence is already a direct experimental signature that the dc bias cancels such detuning terms. However, as will be discussed below, spurious modes can introduce some detuning in the rotating wave approximation at second order. We first focus on the simplest case, the 1-to-1 swap ($n=1$), which exchanges a single memory photon with a single buffer photon. Setting $2eV = \hbar\omega_b - \hbar\omega_a$ brings the Josephson frequency $\omega_J$ to the memory--buffer frequency difference and activates this swap at a rate $g_1$. Probing the buffer by reflectometry (Fig.~\ref{fig:fig2}a), we determine the strength of the 1-to-1 coupling from the dependence of the reflection coefficient $r_b(\omega)$ on SQUID flux (equivalently the effective $E_J$).

When the swap is active, the buffer response develops an absorption dip at $\omega_b$, analogous to optomechanically induced transparency~\cite{weis_optomechanically_2010}. Probe photons resonant with the buffer are converted to the memory and dissipated through its intrinsic losses. As $\hbar g_1=E_J(\Phi)\phi_a\phi_b/2$ increases, this single dip splits into two, corresponding to the hybridized modes, with a separation $2g_1$ reaching up to $2g_1/2\pi \approx 85~\mathrm{MHz}$. The flux $\Phi^c$ that minimizes the reflection amplitude corresponds to critical coupling (Fig.~\ref{fig:fig2}b). Fitting $r_b(\omega)$ with an input--output model that incorporates the bias-voltage noise (full expression in~\cite{supmat}) yields a swap rate $g_1^c/2\pi = 4.8~\mathrm{MHz}$ and a Josephson-frequency linewidth $\sigma_J/2 \pi = 1.5~\mathrm{MHz}$.

This measured rate agrees both with the value $\sqrt{\kappa_b \sigma_J\frac{1}{\sqrt{8\pi}}}= 2\pi \times 5$~MHz expected at critical coupling in the regime $\sigma_J \gg \kappa_a$, and with the independent estimate $g_1(\Phi^c) = E_J(\Phi^c)\phi_a \phi_b / 2\hbar = 2\pi \times 5.9$~MHz obtained from the calibrated $E_J$ and the zero-point fluctuations quoted above. Note that the measured linewidth $\sigma_J$ is well below the buffer decay rate $\kappa_b$, so that bias-voltage noise should not inhibit the engineered dissipation~\cite{aissaoui_cat-qubit-stabilization_2026}.

At small fixed $g_1$, sweeping the bias voltage moves the single dip along the line $\hbar\omega = 2eV + \hbar\omega_a$ (Fig.~\ref{fig:fig2}c). Setting instead $2eV \simeq 2\hbar\omega_a - \hbar\omega_b$ and $2eV \simeq 4\hbar\omega_a - \hbar\omega_b$ activates 2- and 4-photon swaps (shown in Figs.~\ref{fig:fig2}d and \ref{fig:fig2}e respectively). The dips follow $\hbar\omega + 2eV = 2\hbar\omega_a$ (Fig.~\ref{fig:fig2}d) and $\hbar\omega + 2eV = 4\hbar\omega_a$ (Fig.~\ref{fig:fig2}e), confirming the conversion of probe photons into 2 and 4 memory photons. The 4-to-1 swap, in particular, illustrates the ability of the dc-bias approach to reach high-order processes that are difficult to isolate with pumped couplers~\cite{vanselow_dissipating_2026}. Additionally, a 3-photon swap can be activated by setting $2eV \simeq 3\hbar\omega_a - \hbar\omega_b$; however, its effect on the buffer spectroscopy is masked by a bright 1-photon swap between the buffer and the 3/4-wavelength mode of the memory resonator, whose frequency is close to $3\omega_a$.

\begin{figure}[tb]
  \centering
  \includegraphics[width=\columnwidth]{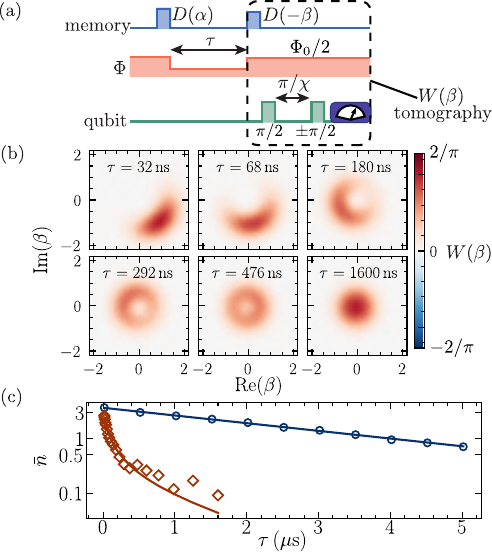}
  \caption{\label{fig:fig3}(a) Pulse sequence: after preparing the memory in a coherent state $|\alpha\rangle$ at flux $\Phi=\Phi_0/2$ (where $E_J$ is negligible), the flux is pulsed to establish a given $E_J$ for a duration $\tau$. Then a memory Wigner tomography is performed at $\Phi=\Phi_0/2$ using the dispersively coupled transmon~\cite{Lutterbach1997,Bertet2002,Vlastakis2013a}.
  (b) Measured Wigner function $W(\beta)$ at increasing $\tau$ for $\alpha=\sqrt{4.1}$ for $E_J/h=2.5~\mathrm{GHz}$ at the fixed voltage $2eV = \hbar(2\omega_a - \omega_b)= h \times 0.53 ~\mathrm{GHz}$ that activates 2-photon loss. The coherent state collapses within ${\sim}50~\mathrm{ns}$ toward a crescent centered on the origin, then relaxes to vacuum over several $\mu$s. (c) Evolution of the average photon number $\bar n(\tau)$ extracted from $W(\beta)$. Blue dots: measurement for $2eV=h \times 0.77~\mathrm{GHz}$, away from any resonant process and $E_J$ kept negligible. Blue line: exponential decay fitting the data with a rate $\kappa_a=(3.1 \pm 0.1~\mu\mathrm{s})^{-1}$. Diamonds: same measurement under the conditions of (b). Orange line: fit to the semiclassical model extracting a two-photon loss rate $\kappa_2/2\pi=0.48~\mathrm{MHz}$. The symbols are larger than the statistical uncertainty.}
\end{figure}

\begin{figure}[tb]
  \centering
  \includegraphics[width=\columnwidth]{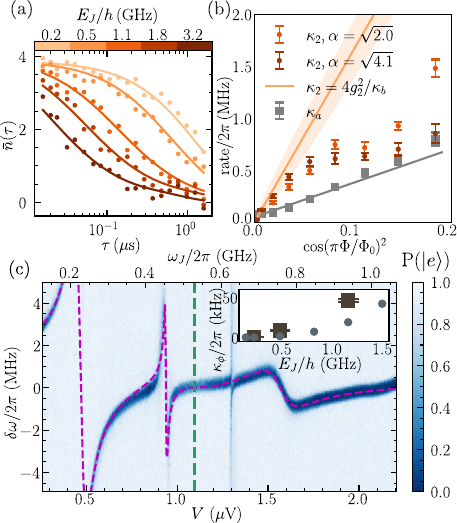}
  \caption{\label{fig:fig4}(a) Dots: extracted average photon number $\bar{n}$ as a function of time $\tau$ under the same 2-photon dissipation conditions as in Fig.~\ref{fig:fig3}b for various values of $E_J$ (colorbar). Lines: fit to a two-mode model~\cite{supmat} that extracts ${g_2}^\mathrm{max}/2\pi =  18.2 \pm 0.5~\mathrm{MHz}$.
(b) Dots with error bars: Effective two-photon dissipation rate $\kappa_2$ as a function of $E_J^2$ extracted from $\bar{n}(\tau)$ using the semiclassical model (Eq.~\ref{eq:semiclassical_model}) for initial coherent amplitudes $\alpha=\sqrt{4.1}$ (red) and $\sqrt{2.0}$ (orange). Solid orange line: expected two-photon dissipation rate $4 g_2^2/\kappa_b$ assuming adiabatic elimination, calculated using ${g_2}^\mathrm{max}$. Squares: $\kappa_a$ extracted from the Wigner function evolution at the origin $W(0, t)$. Solid gray line: linear fit of $\kappa_a$ with $\cos(\pi \Phi/\Phi_0)^2$.
(c) Spectroscopy of the memory as a function of $V$ (equivalently $\omega_J$, top axis) for $E_J/h=0.6$~GHz. The color encodes the probability P($|e\rangle$) to excite the qubit with a selective $\pi$-pulse after driving the memory at $\omega_a+\delta\omega$. Dashed green line: working point $2eV = 2\hbar\omega_a - \hbar\omega_b$. Dashed purple line: fit of the resonant $\delta\omega$ with a perturbative model including spurious modes of the bias line at $227~\mathrm{MHz}$ and $758~\mathrm{MHz}$. At $V=1.3$ $\mu$V the spectroscopy fails due to a parametric process exciting the qubit. Inset: dephasing rate at the working point. Dots: rate computed with $\partial\delta \omega/\partial\omega_J$ and $\sigma_J$, squares: rate extracted from the Wigner functions.}
\end{figure}

We now characterize the engineered two-photon loss of the memory in a ``deflate'' experiment, monitoring the average photon number $\bar n$ stored in the memory as a function of time~\cite{reglade2024quantum}. We determine $\bar n$ from the measured Wigner function $W(\beta)$ of the mode, $\bar n = \int W(\beta)|\beta|^2 \mathrm{d}^2\beta - \tfrac{1}{2}$, measured with the dispersively coupled transmon following the sequence of Fig.~\ref{fig:fig3}a (among the available readouts~\cite{Dassonneville2020}, this one gives the best fidelity in our setup; see \cite{supmat}). As a reference, we start with a working point where $E_J$ is minimal ($\Phi=\Phi_0/2$) and the bias at $2eV= h \times 0.77~\mathrm{GHz}$ so that it is neither $0$ to prevent the SQUID from acting as an inductive coupler, nor at a frequency matching condition that would activate a swap process. In this reference configuration, when the memory is prepared in a coherent state of amplitude $\alpha=\sqrt{4.1}$, we observe that $\bar n(\tau)$ decays exponentially at the intrinsic one-photon rate $\kappa_a = (3.1 \pm 0.1~\mu\mathrm{s})^{-1}$ (Fig.~\ref{fig:fig3}c), as expected for the one-photon loss (here limited by leakage into the flux line).
Setting instead $2eV = 2\hbar\omega_a - \hbar\omega_b$ and activating the swap with a fast flux pulse, which sets the SQUID at $E_J/h=2.5~\mathrm{GHz}$ for a variable duration $\tau$, produces a qualitatively different evolution (Fig.~\ref{fig:fig3}b,c), characteristic of combined one- and two-photon losses with $\kappa_2\gg\kappa_a$. Two-photon loss quickly removes photon pairs at a rate $\simeq |\alpha|^2\kappa_2$, collapsing the coherent state toward the manifold spanned by $|0\rangle$ and $|1\rangle$ within about $50~\mathrm{ns}$; once there, no pairs remain to dissipate and the residual one-photon loss brings the memory to vacuum at a rate $\kappa_a$. The resulting $\bar n(\tau)$ thus shows a fast, non-exponential drop (not a straight line in log scale Fig.~\ref{fig:fig3}c). We explain this evolution with a semiclassical model involving two-photon dissipation (Fig.~\ref{fig:fig3}c), leading to
\begin{equation}
    \bar{n}(t) = \frac{|\alpha|^2 e^{-\kappa_a t}}{1 + \left(1 - e^{-\kappa_a t}\right)2 |\alpha|^2 \kappa_2/\kappa_a}
    \label{eq:semiclassical_model}
\end{equation}
which fits the measured decay with an effective two-photon loss rate $\kappa_2/2\pi=0.48~\mathrm{MHz}$. 
The single photon loss rate $\kappa_a/2\pi=(1.1~\mu\mathrm{s})^{-1}$ is independently extracted from the evolution of the photon number parity $\pi W(0,t)/2$. As shown in Fig.~\ref{fig:fig4}b and discussed below, $\kappa_a$ increases as $E_J^2$. Moreover, under two-photon loss alone, the transient evolution would develop Wigner negativities. Here, they do not appear because strong dephasing induced by the coupling to a spurious mode discussed below prevents their appearance.

We then repeat the deflate experiment over a range of $E_J$, i.e. sweeping $\Phi$, at a fixed initial amplitude $\alpha=\sqrt{4.1}$ (Fig.~\ref{fig:fig4}a). The adiabatic elimination of the buffer turns out to be invalid for the largest $E_J$ values; hence, we fit all measurements jointly to a two-mode model based on Eq.~(\ref{Hamiltonian})~\cite{supmat}, which captures the full memory--buffer dynamics. 
The 2-to-1 photon swap rate reads $g_2 = {g_2}^\mathrm{max}|\cos(\pi\Phi/\Phi_0)|$, and the fitting gives a maximum swap rate ${g_2}^\mathrm{max}/2\pi = 18.2 \pm 0.5~\mathrm{MHz}$.  To characterize the engineered dissipation, we independently extract an effective two-photon loss rate $\kappa_2$ from the fast initial decay of each $\bar n(\tau)$ with the semiclassical model (Eq. \ref{eq:semiclassical_model}), repeating the analysis for two initial amplitudes $\alpha=\sqrt{4.1}$ (red) and $\alpha=\sqrt{2.0}$ (orange) (Fig.~\ref{fig:fig4}b). When the buffer can be adiabatically eliminated, i.e.\ $\kappa_b\gg 8|\alpha|g_2$, the dissipator $\hat L_2=\sqrt{\kappa_2}\hat a^2$ captures the dynamics with $\kappa_2=4g_2^2/\kappa_b$ independent of $\alpha$ and growing quadratically with $E_J$. In this regime, the two amplitudes coincide and match the prediction computed from ${g_2}^\mathrm{max}$. At the largest $E_J$ we could probe, the rates extracted at the two amplitudes separate from each other and from $4g_2^2/\kappa_b$, which is expected when the simple dissipator picture breaks down.

We attribute both the extra memory dephasing and loss to nonlinear processes involving the dc-voltage-biased SQUID and electromagnetic modes in the environment. We attribute the extra loss rate $\kappa_a$ to parasitic 1-photon swap between the memory and modes of the buffer line at $\omega_a\pm\omega_J$\cite{supmat}. The resulting loss rate is $\kappa_a^\mathrm{extra}=\frac{4\phi_a^2}{R_Q}\left(\frac{E_J}\hbar\right)^2\left(\frac{\mathrm{Re}[Z(\omega_a-\omega_J)]}{\omega_a-\omega_J}+\frac{\mathrm{Re}[Z(\omega_a+\omega_J)]}{\omega_a+\omega_J}\right)$, where $Z(\omega)$ is the impedance in parallel to the SQUID. Assuming equal $\mathrm{Re}[Z(\omega)]$ at these two frequencies, the observed slope in Fig.~\ref{fig:fig4}b yields $\mathrm{Re}[Z(\omega_a\pm\omega_J)]={1.8}~\Omega$, which is consistent with our electromagnetic simulations~\cite{supmat}. These frequencies are within the filter stop band. We find that adding 3 periods to the filter should decrease the loss rate by at least 3 orders of magnitude, hence mitigating this problem~\cite{supmat}. The extra dephasing rate growing with $E_J$ (see inset of Fig.~\ref{fig:fig4}c) can be explained by the ac-Stark shift induced by the coupled modes of the environment when they are driven by the dc-voltage-biased SQUID. As a result, the measured memory frequency non-trivially depends on the voltage bias $V$. In Fig.~\ref{fig:fig4}c, the first two splittings correspond to the single- and two-photon driving of a well-identified standing wave mode located between the buffer port and the bias capacitance~\cite{supmat}.  Because $\omega_a$ now depends on the bias, the residual Josephson-frequency noise $\sigma_J/2\pi= 1.5~\mathrm{MHz}$ translates into memory-frequency fluctuations that dephase the memory at a rate $\kappa_\phi=(\partial\delta\omega/\partial\omega_J)\sigma_J$. A perturbative model is able to reproduce the memory spectroscopy (Fig.~\ref{fig:fig4}c).

The memory self-Kerr term $-\hbar K {a^\dagger}^2 a^2/2$ induced by the SQUID is expected to be strongly suppressed by the dynamical averaging that cancels out the other parasitic terms. At zero $E_J$, we measure an upper bound of $K/2\pi<3~\mathrm{kHz}$ for the self-Kerr rate.
Comparing the measured Wigner functions with simulations including various self-Kerr rates $K$ shows that the self-Kerr rate does not visibly increase when $E_J$ rises as expected for dc-voltage-biased junctions~\cite{supmat}. 

In conclusion, we have demonstrated a circuit hosting a high-Q memory resonator coupled to a dc-voltage-biased SQUID, and used it to engineer state-of-the-art single and two-photon loss rates for the memory, as well as to activate up to 4-to-1 photon swaps. The small nonlinearity inherited by the memory and the ability to engineer $n$-photon swaps with $n>2$ could be used in quantum error correction schemes relying on high-order parametric processes, such as the four-legged cat qubit~\cite{cochrane_macroscopically_1999,Leghtas2013,Mirrahimi2014,Ofek2016,grimsmo_quantum_2020,kwon_autonomous_2022,vanselow_dissipating_2026}. Improving the bias circuitry would suppress the spurious resonance that currently limits the memory coherence at large swap rates. A bias setup based on the dc-Josephson effect~\cite{smirr2025tunable,esmaeili_phase-stable_2026} or on injection locking~\cite{aissaoui_cat-qubit-stabilization_2026,danner_quantum_2025} could eliminate the dephasing due to bias-voltage noise, paving the way for versatile parametric couplers with large swap rates, compatible with high-Q modes. Finally, adding a two-photon drive on such a device, either using the dc-biased junction or by pumping a dedicated nonlinear coupler, would stabilize the memory state and turn it into a cat qubit.

\begin{acknowledgments}
We thank T. Aissaoui, J. Ankerhold, A. Clerk, A. Guimbal, M. Hofheinz, B. Kubala, C. Padurariu, and A. Sarlette for useful discussions. This work received government funding administered by the French National Research Agency (ANR) under the France 2030 program, grant numbers ANR-22-PETQ-0003, ANR-22-PETQ-0006, as well as iDemo grant UsineAChats from BPI. A. B. acknowledges support from the “Fondation CFM pour la Recherche". Claude Sonnet 4.6 and Claude Opus 5 by Anthropic were used to improve the readability of the text.
\end{acknowledgments}

\nocite{yang_hot_2025,peugeot_quantum_2020,di2010simple,Cahill1969,flurin2014josephson,schuster2007resolving,martel2025influence,gasser1986residual,guilmin2025dynamiqs,jax2018github}

\bibliography{biblio}

\end{document}


\preprint{APS/123-QED}

\title{
Supplemental material for\\
"Engineering multi-photon dissipation with a dc-voltage-biased Josephson junction"}

\author{Marco Paradina}
\affiliation{Ecole Normale Sup\'erieure de Lyon,  CNRS, Laboratoire de Physique, F-69342 Lyon, France}
\affiliation{Alice \& Bob, 53 Bd du G\'en\'eral Martial Valin, 75015 Paris, France}

\author{Ambroise Peugeot}%
\email[contact: ]{ambroise.peugeot@ens-lyon.fr}
\affiliation{Ecole Normale Sup\'erieure de Lyon,  CNRS, Laboratoire de Physique, F-69342 Lyon, France}

\author{Roberto Negrin}
\affiliation{Alice \& Bob, 53 Bd du G\'en\'eral Martial Valin, 75015 Paris, France}
\affiliation{Laboratoire de Physique de l’École Normale Supérieure, Mines Paris, Inria, CNRS, ENS-PSL, Centre Automatique et Systèmes (CAS),
Sorbonne Université, PSL Research University, Paris, France}

\author{Oscar Novat}%
\affiliation{Ecole Normale Sup\'erieure de Lyon,  CNRS, Laboratoire de Physique, F-69342 Lyon, France}

\author{Tristan Villain}%
\affiliation{Ecole Normale Sup\'erieure de Lyon,  CNRS, Laboratoire de Physique, F-69342 Lyon, France}

\author{Anil Murani}
\affiliation{Alice \& Bob, 53 Bd du G\'en\'eral Martial Valin, 75015 Paris, France}

\author{Jean-Loup Ville}
\affiliation{Alice \& Bob, 53 Bd du G\'en\'eral Martial Valin, 75015 Paris, France}

\author{Sébastien Jezouin}
\affiliation{Alice \& Bob, 53 Bd du G\'en\'eral Martial Valin, 75015 Paris, France}

\author{Raphaël Lescanne}
\affiliation{Alice \& Bob, 53 Bd du G\'en\'eral Martial Valin, 75015 Paris, France}

\author{Audrey Bienfait}%
\affiliation{Ecole Normale Sup\'erieure de Lyon,  CNRS, Laboratoire de Physique, F-69342 Lyon, France}

\author{Benjamin Huard}%
\affiliation{Ecole Normale Sup\'erieure de Lyon,  CNRS, Laboratoire de Physique, F-69342 Lyon, France}
 
\date{\today}

\maketitle

\tableofcontents

\renewcommand{\thefigure}{S\arabic{figure}}
\setcounter{figure}{0}
\renewcommand{\thetable}{S\arabic{table}}
\setcounter{table}{0}
\renewcommand{\theequation}{S\arabic{equation}}
\setcounter{equation}{0}

\section{Experimental setup and device parameters}

\begin{figure}[h!]
    \centering
    \includegraphics[width=0.7\linewidth]{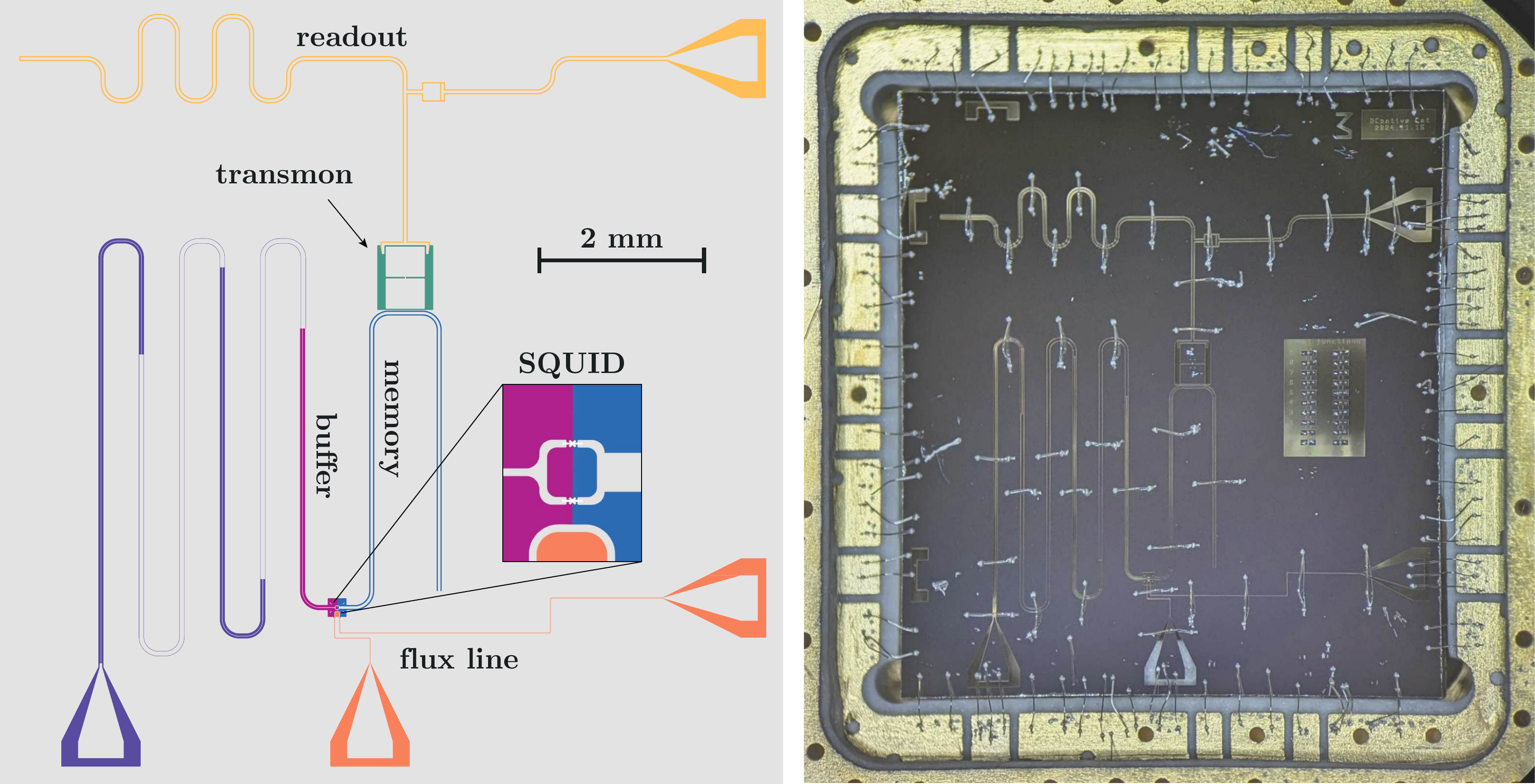}
    \caption{Left panel: Drawing of the device. An inset shows a schematic of the SQUID position and its flux line. Right panel: Optical image of the device in its sample holder.}
    \label{fig:device}
\end{figure}

\begin{sidewaysfigure}
    \centering
    \includegraphics[width=\textwidth]{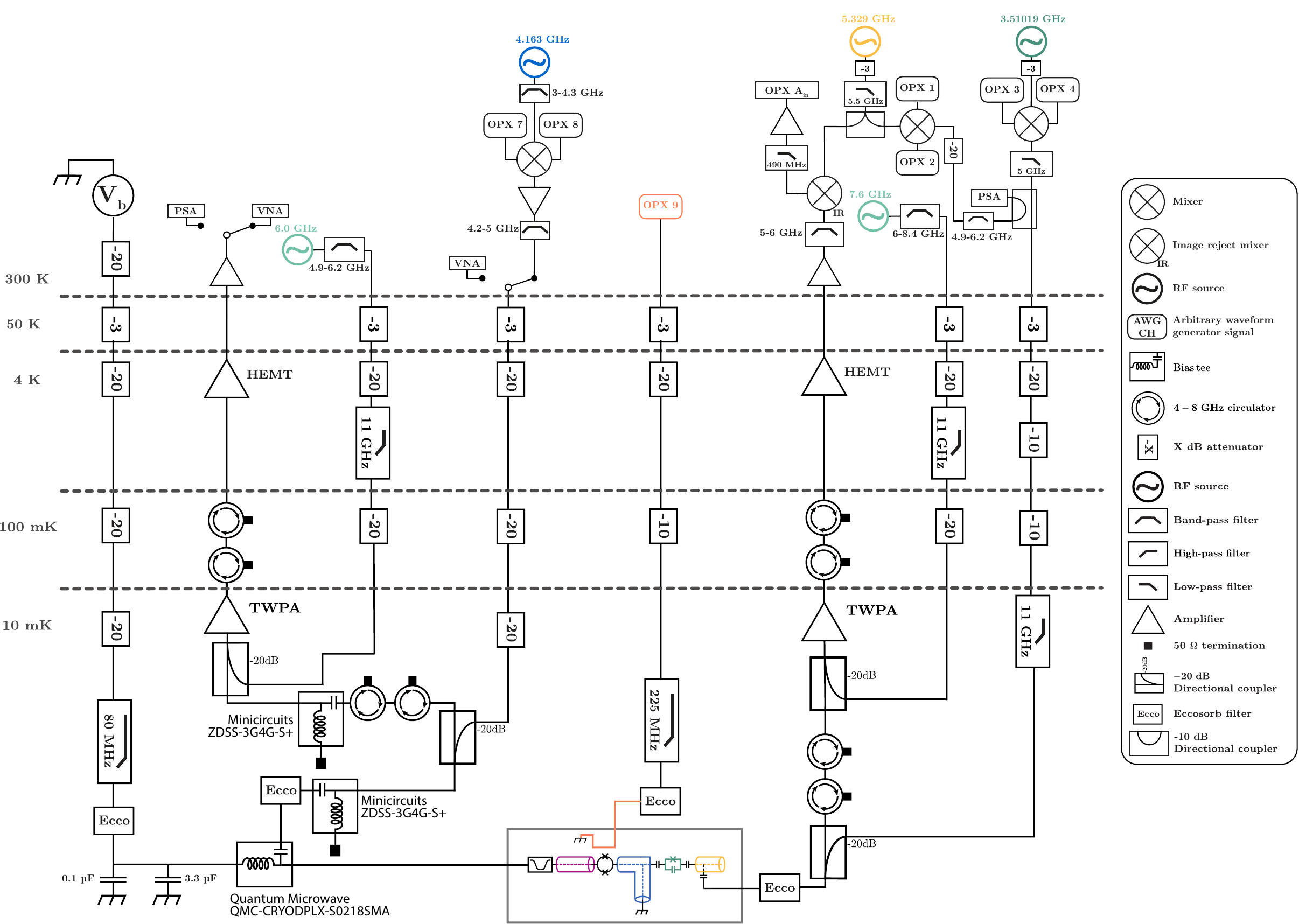}
    \caption{Wiring diagram of the experiment. The OPX boxes represent DAC (rounded) or ADC (square) channels of a single OPX+ instrument by Quantum Machines. The local oscillators are frequency converted using an IQ mixer and I and Q modulating signals coming out of the OPX+. Depending on the experiment, the buffer input and output lines are connected to either a Vector Network Analyzer (VNA), a Power Spectrum Analyzer (PSA) or a pulsed measurement and control setup. Room temperature isolators are not shown.}
    \label{fig:cabling}
\end{sidewaysfigure}

The drawing of the device and its picture are shown in Fig.~\ref{fig:device}. All lines are coplanar waveguides. Their various gap size and central line widths are the parameters used to control their characteristic impedance, in particular in the stepped impedance filter at the input of the buffer line. All the device parameters can be found in Table~\ref{tab:device_parameters}.

The scheme of the measurement setup is shown in Fig.~\ref{fig:cabling}. The dc voltage bias is implemented by voltage biasing a coaxial cable (with a graphite-coated dielectric insulator) at room temperature using a Bilt module at a bias voltage $V_b$. The filtering at base temperature is shown in the figure. The fast flux line is directly connected to a DAC channel of the OPX+ (by Quantum Machines). The pulsed measurements are performed using standard I-Q mixing and demodulating, therefore the LO frequencies are offset from the actual drive frequencies received at the device position.

\newcommand{\prow}[3]{#1 & #2 & #3 \\}
 
\begin{table*}[htbp]
\caption{Parameters of the device and the associated determination method.}
\label{tab:device_parameters}
\begin{ruledtabular}
\begin{tabular}{lcl}
\textbf{Parameter} & \textbf{Value} & \textbf{Method of determination} \\
\hline
\prow{Memory frequency $\omega_a/2\pi$}{$4.0429 \rm \, GHz$}{Memory spectroscopy using the qubit (3-tone)}
\prow{Buffer frequency $\omega_b/2\pi$}{$ 7.56 \rm \, GHz$}{Resonator spectroscopy}
\prow{Transmon frequency $\omega_q/2\pi$}{$3.412 \rm \, 
GHz$}{Qubit (2-tone) spectroscopy}
\prow{Readout resonator frequency $\omega_{\mathrm{ro}}/2\pi$}{$5.379 \rm \, GHz$}{Resonator spectroscopy}
\hline
\prow{Maximum SQUID Josephson energy $E_J^{\mathrm{max}}/h$}{$ 13\pm2 \rm \, GHz$}{From the resonators frequencies modulation with flux}
\prow{SQUID asymmetry $\epsilon$}{$0.016\pm 0.005$}{From the power emitted by the SQUID as a function of flux}
\prow{Memory zero-point phase fluctuation $\phi_a$}{0.18}{Design}
\prow{Buffer zero-point phase fluctuation $\phi_b$}{0.21}{Design}
\prow{Maximum two-to-one coupling rate $g_2^{\mathrm{max}}/2\pi$}{$18.2 \pm 0.5 \rm \, MHz$}{Fit of photon number decay (Sec.~\ref{sec:deflate_numerics})}
\prow{Voltage noise $\sigma_{\mathrm{J}}/2\pi$}{$1.5 \rm \, MHz \, (3.1~nV)$}{Josephson emission spectroscopy}
\hline
\prow{Memory bare self-Kerr $K^{\rm bare}/2\pi$}{$<3 \rm \, kHz$}{Displacing to vacuum~\cite{yang_hot_2025}}
\prow{Memory self-Kerr under 2-to-1 process $K^{\rm on}/2\pi$}{$<10 \rm \, kHz$}{Fits of Wigner function dynamics}
\prow{Transmon to memory cross-Kerr $\chi_{qa}/2\pi$}{$1.51 \rm \, MHz$}{Ramsey interferometry with populated memory}
\prow{Transmon to readout cross-Kerr $\chi_{q\mathrm{ro}}/2\pi$}{$2.5 \rm \, MHz$}{Readout spectroscopy}
\hline
\prow{Memory single-photon relaxation time $T_1^{\rm mem}$}{$3.1 \pm 0.1  \rm \, \mu s$ }{Wigner tomography}
\prow{Buffer loss rate $\kappa_b/2\pi$}{$85 \pm 5 \rm \, MHz$}{Buffer spectroscopy}
\prow{Readout resonator linewidth $\kappa_r/2\pi$}{$1.13 \rm \, MHz$}{Readout spectroscopy}
\prow{Transmon relaxation time $T_1^q$}{$103 \pm 2 \rm \, \mu s$}{Standard decay measurement}
\prow{Transmon Ramsey coherence time $T_2^{*}$}{$13.3 \pm 0.2 \rm \, \mu s$}{Ramsey interferometry}
\prow{Transmon echo coherence time $T_2^{\mathrm{echo}}$}{$23.3 \pm 0.3 \rm \, \mu s$}{Ramsey interferometry}
\hline  
\prow{Memory thermal population $n_{\mathrm{th}}^{a}$}{$0.005$}{Memory vacuum detector}
\prow{Transmon thermal population $n_{\mathrm{th}}^{q}$}{$0.008$}{Transmon readout}
\end{tabular}
\end{ruledtabular}
\end{table*}

\section{Resonator spectroscopies}

\begin{figure}[htbp]
    \centering
    \includegraphics[width=0.9\textwidth]{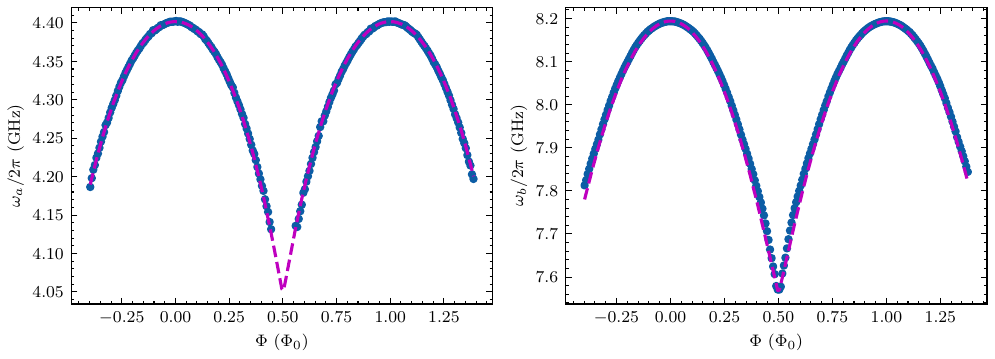}
    \caption{
   Frequency of the memory (left panel) and buffer (right panel) resonators as a function of the flux $\Phi$ threading the SQUID loop. Dots: experimental data extracted from microwave reflection spectroscopy, dashed line: fits to a model of the circuit.}
    \label{fig:resonator_spectroscopies}
\end{figure}

In the absence of a bias voltage ($V=0$), the SQUID behaves as a flux-tunable inductive
element. Because the resonators have a $\lambda/4$ geometry (Fig.~\ref{fig:device}), the SQUID is located at a voltage antinode of both the buffer and the memory. It therefore affects the boundary conditions and, consequently, their resonance frequencies. Tuning the SQUID inductance allows us to tune both resonator frequencies. Figure~\ref{fig:resonator_spectroscopies} shows the buffer and memory direct spectroscopies measured in reflection through the buffer transmission line while sweeping the flux line bias current, converted into the flux $\Phi$ through the SQUID loop.

The measurement is performed with a Vector Network Analyzer (VNA) connected to the buffer transmission
line. We record the phase of the reflected signal $r_b(\omega)$ as a function of the probe
frequency $\omega$ and the flux bias. The buffer, an overcoupled resonator, shows up as a $2\pi$
phase drop, whereas the memory, an undercoupled resonator, appears only as a small phase
wiggle. By tracking the positions of these features as a function of the flux bias, we
extract the dependence of the resonator frequencies on the flux through the SQUID in Fig.~\ref{fig:resonator_spectroscopies}.
The buffer frequency is extracted by fitting the phase response to $\arg(r_b)\left(\omega\right)=2\arctan \left(2\frac{\omega-\omega_b}{\kappa_b}\right)$. The memory frequency is harder to fit over the whole frequency span. We thus extract it numerically as the maxima of the gradient of the
reflected phase with respect to frequency, $|\partial \arg(r_b / \partial \omega|$.

We reproduce the resonator frequencies by modeling the phase response $\arg(r_b(\omega))$ using several fit parameters: the characteristic impedance and length of the transmission line segments, the SQUID inductance and the stray capacitance. The asymmetry between the two junctions of the SQUID is too small to have visible impact on the spectroscopies, thus we do not consider it in the model. This asymmetry is estimated and discussed in more detail in Sec.~\ref{sec:squid_asymmetry}.

\section{Details on the bias setup and the voltage noise}
\subsection{Low-noise bias voltage setup}
We use an attenuated coaxial line combined with low-pass filtering at the mixing chamber stage to apply a bias voltage across the SQUID with sufficiently low voltage noise (Fig.~\ref{fig:cabling}).

The attenuated line effectively behaves as a $R\simeq$ 50 $\Omega$ bias resistor in parallel with the SQUID. We add a $C=$ 3.4 $\mu$F capacitance in parallel in order to filter out voltage noise reaching the junction. Note that in practice, we use the parallel combination of a 3.3 $\mu$F capacitor and a 100 nF capacitor, in order to minimize the impact of their stray inductance. This low-pass filter has a cutoff frequency $1/RC\simeq2\pi\times$ 1 kHz, so that we can sweep the bias across the SQUID in a fraction of a second.

\subsection{Calibrating voltage divider and offset}
\label{sec:divider_calibration}

To bias the junction, we set a voltage $V_\mathrm{b}$ on
the voltage source at room temperature. The voltage $V$ bias across the SQUID
is scaled down by the voltage divider formed by the
series resistance of the heavily attenuated bias line, and it is shifted by
thermoelectric offsets of a few mV. This is captured by the relation
\begin{equation}
V = \frac{V_\mathrm{b} - V_\mathrm{offset}}{D}
\label{eq:divider}
\end{equation}
where $D$ is the voltage divider ratio and $V_\mathrm{offset}$ is the offset. Neither of these quantities is known a priori with sufficient accuracy, so we calibrate them in situ.

The calibration relies on Josephson emission. When the junction is biased such that
$2eV = \hbar\omega$ with $\omega$ lying within the buffer band, the junction emits photons at the Josephson frequency $\omega_J = 2eV/\hbar$ out of
the buffer port~\cite{Hofheinz2011, Rolland2019}.  

\begin{figure}[h]
    \centering
    \includegraphics[width=\textwidth]{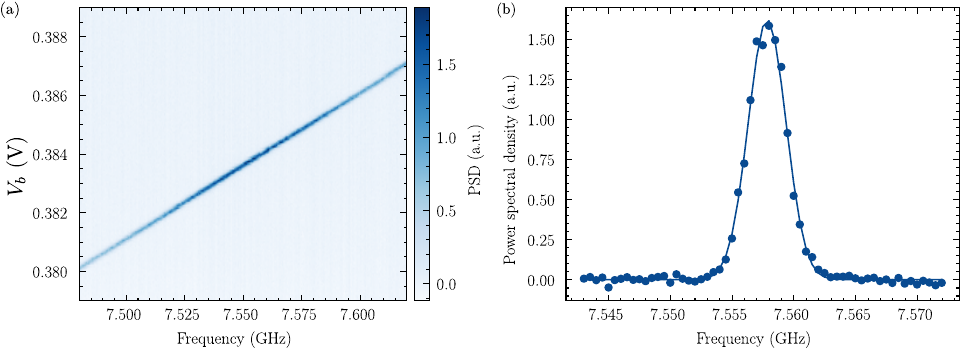}
\caption{(a) Power spectral density (PSD) of the radiation emitted by the SQUID as a function of the bias voltage $V_b$ and of frequency. (b) PSD measured a fixed bias $V=15.629$~$\mu$V such that $\omega_J/2\pi=7.558$~GHz. Blue dots are data, the continuous line is a fit to a Gaussian function with $\sigma=1.5$~MHz.}
    \label{fig:voltage_bias_calibration}
\end{figure}

We therefore sweep
$V_\mathrm{b}$ and measure the power
spectral density at the output of the buffer line (Fig.~\ref{fig:cabling}) with a
spectrum analyzer. This yields an emission map (Fig.~\ref{fig:voltage_bias_calibration}a) as a function of acquisition
frequency and bias voltage. 

At each bias point, the emission spectrum displays a
peak (Fig.~\ref{fig:voltage_bias_calibration}b) centered at $\omega_J$ with a Gaussian shape~\cite{Rolland2019, peugeot_quantum_2020}. 

Fitting each spectrum with a
Gaussian gives the center frequency, and hence $\omega_J$, as a function of
$V_\mathrm{b}$. The fitted centers fall on a line,

\begin{equation}
\omega_J(V_\mathrm{b}) = \frac{2e}{\hbar}\,\frac{V_\mathrm{b} - V_\mathrm{off}}{D},
\end{equation}

whose slope and intercept directly yield the divider ratio and the offset.

All bias voltages $V$ quoted in this work are junction voltages obtained through
Eq.~\eqref{eq:divider}.

\subsection{Characterizing the voltage noise}
\label{sec:voltage_noise}

Bias voltage noise $S_{VV}(\omega)$ induces jitter of the Josephson frequency $\omega_J$, with a power spectral density $S_{\omega_J\omega_J}(\omega) = 4e^2/\hbar^2 \times S_{VV}(\omega)$, which is approximately flat up to the cut-off frequency $1/RC$ of the bias line imposed by the shunt capacitor. In the limit where the cut-off frequency is much smaller than $S_{\omega_J\omega_J}(0)$~\cite{di2010simple}, the frequency jitter can be considered as a slow modulation of $\omega_J$, which follows faithfully the instantaneous value of $2eV/\hbar$. Under these conditions, the emission spectrum of the junction is a Gaussian function with width parameter $\sigma_J = 2e / \hbar \sigma_V$, reflecting the Gaussian nature of the voltage noise.

We can characterize the amplitude of the voltage noise across the junction by fitting the emission spectra to a Gaussian function (Fig.~\ref{fig:voltage_bias_calibration}b). We find widths $\sigma_J / 2\pi$ varying in time between $1.1$ and $1.5~\mathrm{MHz}$, corresponding to an RMS voltage noise  $\sigma_V$ between $2.3~\mathrm{nV}$ and $3.1~\mathrm{nV}$.

As the linewidth $\sigma_J$ remains far below the buffer loss rate $\kappa_b$, it does not affect the engineered dissipation rates~\cite{aissaoui_cat-qubit-stabilization_2026}.

We note that the emission peak becomes sharper when we turn off the pulse tube cryocooler, with a linewidth down to about 300 kHz corresponding to $\sigma_V = 620$~pV. The extra voltage noise we measure with the pulse tube turned on could be explained either by an elevated electron temperature of the bias resistor, or due to triboelectricity or piezo-electrical effects.

\subsection{SQUID asymmetry}
The total Josephson energy of the SQUID reads 

\begin{equation}
E_J(\Phi)=E_J^{\mathrm{max}}\sqrt{\cos^2(\pi\Phi/\Phi_0)+\epsilon^2\sin^2(\pi\Phi/\Phi_0)},\label{eq:EJofphi_asym}
\end{equation}
where $E_J^{\mathrm{max}}$ is the sum of the Josephson energies of the two junctions, and the asymmetry parameter $\epsilon$ is defined as their difference normalized by $E_J^{\mathrm{max}}$. Even though the two junctions are designed to be nominally identical, in practice they have slightly different $E_J$ values due to nanofabrication uncertainties, such that we expect $\epsilon$ to be typically of a few percents.

\begin{figure}[h]
    \centering
    \includegraphics[width=0.45\textwidth]{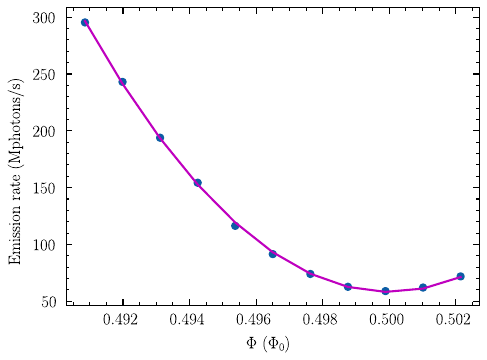}
\caption{Photon emission rate as a function of the flux through the SQUID. Dots: experimental data, solid line: fit using square of Eq.~(\ref{eq:EJofphi_asym}) yielding an asymmetry parameter $\epsilon=0.016\pm0.005$ between the junctions of the SQUID. The emission rate is calibrated following Ref.~\cite{peugeot_quantum_2020}.
    \label{fig:asymmetry}}
\end{figure}

We experimentally determine $\epsilon$ in our device by measuring the microwave power emitted by the SQUID in the AC-Josephson effect regime, which scales like $E_J(\Phi)^2$ and would go down to zero at $\Phi=\Phi_0/2$ in the case of a perfectly symmetric SQUID. We tune $2eV$ close to $\hbar\omega_b$ and sweep $\Phi$ around $\Phi_0/2$, measuring the power emitted by the SQUID at frequency $\omega_b$. By fitting the resulting flux dependence with the square of Eq.~(\ref{eq:EJofphi_asym}) (line in Fig.~\ref{fig:asymmetry}), we obtain an asymmetry parameter $\epsilon=0.016\pm0.005$. The uncertainty on $\epsilon$ is estimated by repeating this measurement at different bias voltages and measuring the emission power at the corresponding frequencies.

\label{sec:squid_asymmetry}

\section{Characterization of the buffer mode and of the bias line spurious mode}
Measuring the emission spectra of the SQUID versus voltage bias also allows us to characterize the modes in the environment of the SQUID. We explain here how we apply this method to the buffer resonator, and to a low-frequency mode that we attribute to a standing wave of the bias line.
\subsection{Buffer impedance}
\label{sec:buffer_impedance}

The Josephson emission also provides an in-situ measurement of the impedance
in parallel to the SQUID, and hence of the buffer mode profile. In the perturbative
regime of small $E_J$, the rate at which Cooper pair tunneling emits photons
into the circuit is given by $P(E)$ theory~\cite{ingold_charge_1992, Hofheinz2011}
as $\Gamma_{\rm ph} = \frac{\pi E_J^2}{2\hbar}\, P(2eV)$, where the probability density of releasing the tunneling energy into a
single photon reduces to
\begin{equation}
P(2eV) \simeq \frac{2\,\mathrm{Re}\!\left[Z(\omega_J)\right]}{R_Q\,\hbar\omega_J},
\label{eq:PofE}
\end{equation}
with $\omega_J = 2eV/\hbar$, $Z(\omega)$ the circuit impedance in parallel to the SQUID and $R_Q = h/4e^2$ the resistance quantum. The
power emitted at the Josephson frequency, $P_\mathrm{em} = \hbar\omega_J\,\Gamma_{\rm ph}
= \frac{\pi E_J^2}{\hbar R_Q}\,\mathrm{Re}\!\left[Z(\omega_J)\right]$, is therefore directly proportional to the real part of the impedance at the emission frequency.

\begin{figure}[h]
    \centering
    \includegraphics[width=0.5\textwidth]{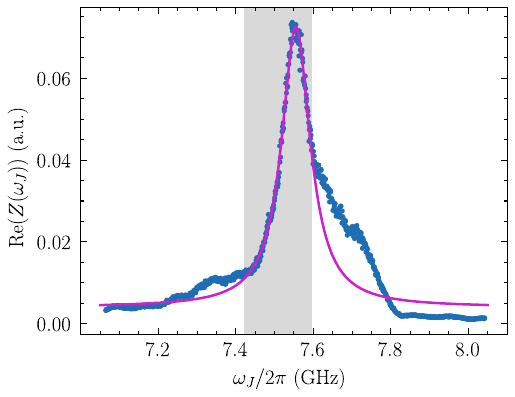}
    \caption{Real part of the impedance in parallel to the SQUID $\mathrm{Re}[Z(\omega_J)]$ close to the buffer frequency. Blue dots: data extracted from the area under the emission peak. Magenta line: fit to a Lorentzian function yielding $\kappa_b/ 2 \pi = 85 \pm 5 \, \rm MHz$. Only the data points in the gray-colored $\omega_J$ range are used for the fit.}
    \label{fig:buffer_impedance_profile}
\end{figure}

We exploit this proportionality by sweeping the bias voltage across the buffer
band and recording the emission spectrum at each bias point, as in
Sec.~\ref{sec:divider_calibration}. At each voltage, we fit the emission peak
and extract its area, which measures the total emitted power at $\omega_J$.
Plotting this area against $\omega_J$ then traces out
$\mathrm{Re}\!\left[Z(\omega)\right]$ in parallel to the SQUID, up to a global
scale factor set by $E_J$ and the frequency-dependent gain of the
amplification chain, which we do not calibrate independently. The resulting profile is shown in Fig.~\ref{fig:buffer_impedance_profile}.

\subsection{Standing-wave mode of the bias line}
\label{sec:sw_mode}

\begin{figure}[h]
    \centering
    \includegraphics[width=0.9\textwidth]{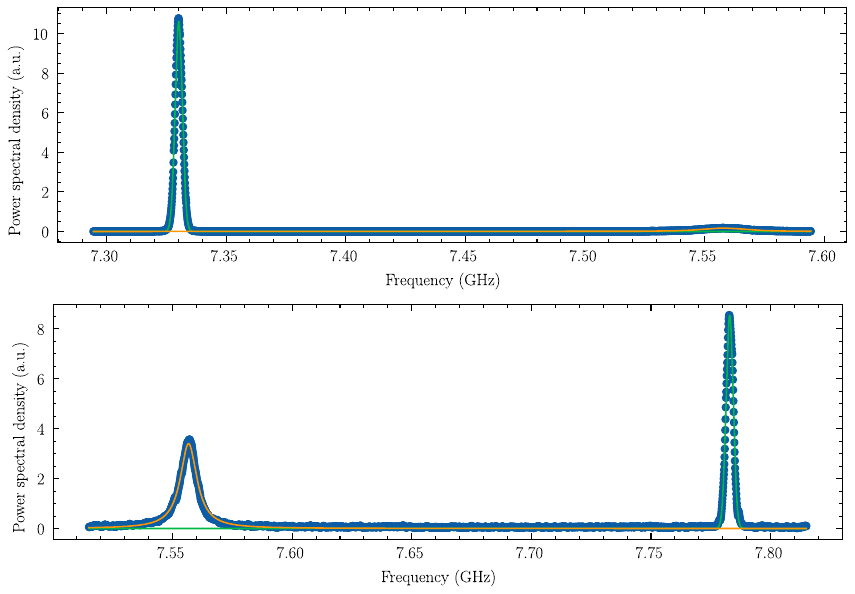}
    \caption{PSD of the radiation emitted by the SQUID with the bias voltage set to $2eV=\hbar(\omega_b-\omega_c)$ (upper panel) and $2eV=\hbar(\omega_b+\omega_c)$ (lower panel). Blue dots: experimental data, green line: fit to a Gaussian line shape with width $\sigma=1.5$~MHz, orange line: fit to a Lorentzian profile with full width at half-maximum $\kappa_c/2\pi=14$~MHz.}
    \label{fig:standing_wave_mode_emission}
\end{figure}

Enlarging the span of the Josephson emission map of Sec.~\ref{sec:divider_calibration} reveals a spurious mode of the circuit. In addition to the expected emission peak at the buffer frequency, following $\hbar\omega = 2eV$, we observe two fainter lines running parallel to it, symmetrically displaced on either side by the same frequency offset (not shown). These sidebands are the signature of parametric processes involving not only the buffer but also an additional mode $c$, at a frequency $\omega_c$ given by the offset of the sidebands.

To confirm this interpretation and characterize the mode $c$, we probe the two
processes separately by setting the bias voltage to the corresponding
energy-matching conditions. Setting $2eV = \hbar(\omega_b - \omega_c)$, we
observe an emission peak at $\omega_J = \omega_b - \omega_c$
(Fig.~\ref{fig:standing_wave_mode_emission}): the energy $2eV$ of a tunneling Cooper pair is
completed by the absorption of one photon from mode $c$ to emit a single
photon into the buffer. This anti-Stokes process is visible because mode $c$
is populated at thermal equilibrium, as expected from its frequency $\hbar\omega_c/k_B \approx 11~\mathrm{mK}$. Setting instead $2eV = \hbar(\omega_b + \omega_c)$, we
observe a peak at $\omega_J = \omega_b + \omega_c$
(Fig.~\ref{fig:standing_wave_mode_emission}), corresponding to the Stokes process in which
the tunneling energy is spent for the correlated emission of one buffer
photon and one photon in mode $c$~\cite{Hofheinz2011, Peugeot2021}.
Fitting the emission peaks yields $\omega_c/2\pi = 227$~MHz and a linewidth
$\kappa_c/2\pi = 14$~MHz.

We identify mode $c$ as a standing wave of the bias line, confined between the capacitor of the bias tee on the left of the buffer in Fig.~\ref{fig:cabling} and the buffer port on the chip. This hypothesis is supported by
the fact that reducing the electrical length of this section of line between cooldowns shifts $\omega_c$ upwards.

\section{Measuring the mean photon number}
\label{sec:measuring_memory}

Several techniques can be used to measure the mean photon number in the memory. After considering populated Ramsey interferometry~\cite{Dassonneville2020}, direct readout through the buffer~\cite{reglade2024quantum}, and other techniques, we decided to infer the photon number from the Wigner function directly. Indeed, direct Wigner tomography is compatible with the relatively short memory lifetime. That is because all pulses are unconditional with respect to the qubit state, and they can be made short compared to $T_1^{\rm mem}$.

In practice, the duration of the pulses in the Wigner tomography sequence (Fig.~\ref{main-fig:fig3}) are
\begin{itemize}
    \item $D(-\beta)$ displacement pulse $20 \rm \, ns$ 
    \item $\pi/2$ pulse: $36 \rm \, ns$
    \item $\pi/\chi$ wait time: $328 \rm \, ns$.
\end{itemize}

The mean photon number of the memory is obtained as~\cite{Cahill1969}
\begin{equation}
\label{eq:wigner_integral}
    \bar n = \int W(\beta)\,|\beta|^2\, \mathrm{d}^2\beta - \frac{1}{2} .
\end{equation}

The memory population $\bar{n}$ extracted with Eq.~\eqref{eq:wigner_integral} needs to be corrected for the following two parasitic effects.

\paragraph{Wigner function axes normalization}
To calibrate the axes of the Wigner function, we measure the Wigner function of the memory in its equilibrium state, which is very close to the vacuum state. We scale the axes of the Wigner function to match its width to a Gaussian lineshape with $\sigma=0.5$ as expected for the vacuum state. 
Nonetheless, we observe that this calibration is not valid anymore as soon as we turn on the 2-photon losses. We attribute this to a possible effect of the dc-voltage-biased SQUID and its flux pulsing onto the transmon qubit that is used to perform the Wigner tomography. We thus rescale the axes of the Wigner functions by $c(\Phi)$ for every flux bias value $\Phi$, using the vacuum state as a reference.

This calibration was performed three months after the deflate experiment. We observed that there had been an additional drift in the experiment, resulting in a systematic rescaling of $c(\Phi)$ in our final data treatment as detailed in Sec.~\ref{sec:deflate_numerics}.

\paragraph{Effect of memory $T_1^{\rm mem}$ }

The memory lifetime $T_1^{\rm mem} = 3.1 \, \rm \mu s$ is comparable to the duration of the Wigner tomography pulse sequence $t_{\rm Wigner} = 420 \rm \, ns$. This means that the memory significantly decays during Wigner tomography, leading to an underestimation of the memory population $\bar{n}(\tau)$. We simulated the memory decay during the pulse sequence, and rescaled the value of $\bar{n}(\tau)$ to account for it.

\section{Fitting the 1-to-1 conversion data}

Including a 1-photon conversion term $\hbar g_1 e^{i\delta\omega_Jt}a^\dagger b+h.c.$ in the circuit Hamiltonian, we find that the reflection coefficient is~\cite{flurin2014josephson}:
\begin{equation}
\label{eq:g1_reflection1}
r_b(\Delta,\delta\omega_J)=\frac{(\Delta-\delta\omega_J+i\kappa_a/2)(\Delta-i\kappa_b/2)-|g_1|^2}{(\Delta-\delta\omega_J+i\kappa_a/2)(\Delta+i\kappa_b/2)-|g_1|^2}
\end{equation}

In our experiment, the bias voltage is not static but fluctuates randomly, on a timescale much slower than the resonators lifetime. We take this into account by convolving Eq.~\eqref{eq:g1_reflection1} with a Gaussian distribution of width $\sigma_J$. Using Claude Sonnet 4.6 by Anthropic, we find:
\begin{equation}
\label{eq:g1_reflection}
\langle r_b(\Delta)\rangle=\frac{(\Delta-i\kappa_b/2)}{(\Delta+i\kappa_b/2)}-\sqrt{\frac{\pi}2}\frac{\kappa_b|g_1|^2}{\sigma_J(\Delta+i\kappa_b/2)^2}w\left(\frac{z_0}{\sqrt{2}\sigma_J}\right)
\end{equation}
with $w(z)$ the Faddeeva function and $z_0=\Delta +i\kappa_a/2-|g_1|^2/(\Delta+i\kappa_b/2)$.

\begin{figure}[h]
    \centering
    \includegraphics[width=0.9\textwidth]{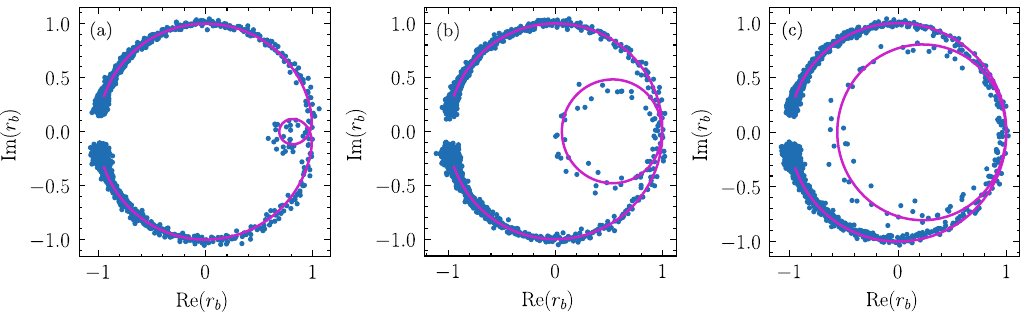}
\caption{Buffer reflection spectroscopy measured with a finite $g_1$ swap rate corresponding to a flux value such that $\cos(\pi\Phi/\Phi_0)$ is 0 (a), 0.058 (b), and 0.110 (c). Blue points: data, solid line: fit to Eq.~\eqref{eq:g1_reflection}. We find $g_1/2\pi = 3 \pm 1 \rm \, MHz$,  $g_1/2\pi = 6 \pm 1 \rm \, MHz$ and $g_1/2\pi =10 \pm 1 \rm \, MHz$ for panels (a), (b), (c) respectively.}
    \label{fig:VNA_buffer_fit}
\end{figure}

\section{Losses and dephasing of the memory resonator}

\subsection{Limitation to the memory lifetime}
The memory resonator has a relatively short lifetime of 3.1~$\mu$s corresponding to a total quality factor of about $Q_{tot}=8\times10^4$, well below the state of the art for tantalum 2D resonators where $Q_{tot}=10^6$ can be achieved. We discuss here the plausible causes of this low quality factor and how to remedy it in a future iteration of the experiment.

We define four effective loss channels for the memory resonator, so that the total inverse quality factor is:
\begin{equation}
    {Q_{tot}}^{-1}={Q_{buff}}^{-1}+{Q_{ro}}^{-1}+{Q_{flux}}^{-1}+{Q_{int}}^{-1}
\end{equation}
Three loss channels correspond to leakage out of the chip towards microwave lines, respectively the buffer line, the qubit readout line, and the flux line, with contribution ${Q_{buff}}^{-1}$, ${Q_{ro}}^{-1}$ and ${Q_{flux}}^{-1}$ respectively. The last one (internal losses with ${Q_{int}}^{-1}$) corresponds to dissipation inside the resonator, which could be due e.g. to suboptimal nanofabrication yielding larger than state-of-the-art dielectric losses. We want to ascertain which is the dominating loss channel.

\begin{figure}[h]
    \centering
    \includegraphics[width=0.9\textwidth]{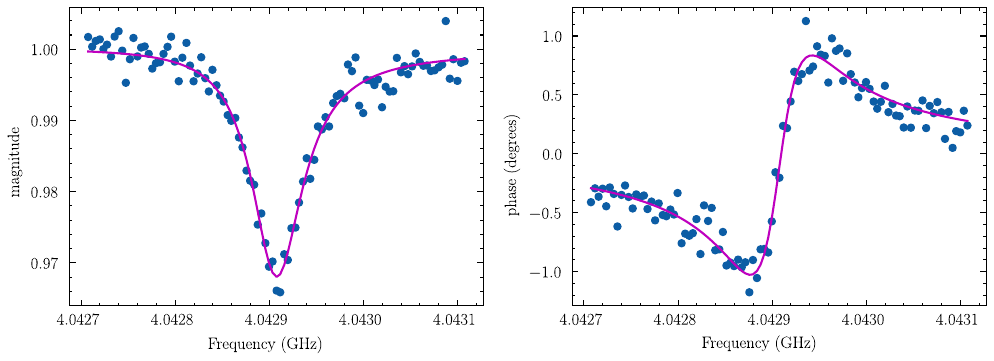}
\caption{Magnitude (left panel) and phase (right panel) of the complex reflection coefficient on the memory resonator as a function of frequency, measured from the buffer port. Blue points: experimental data, purple line: fit to a model including a finite Fano resonance to account for the asymmetric lineshape. We find $Q_{buff}=4\times10^6$ and $Q_{tot}=8\times10^4$.}
    \label{fig:VNA_memory_fit}
\end{figure}

Using the VNA, we first perform a microwave reflection measurement close to the memory frequency through the buffer line (Fig.~\ref{fig:VNA_memory_fit}). We fit the complex reflection coefficient and find that it is deep in the undercoupled regime, with the coupling to the buffer line $Q_{buff} =4\times10^6\gg Q_{tot}$, validating that leakage to the buffer line does not limit the memory lifetime.

The same measurement through the qubit readout port yields no amplitude or phase modulation at all across $\omega_a$, indicating that the coupling to this line is even smaller than to the buffer line: $Q_{ro} \gg4\times10^6$. We can thus also exclude leakage to the qubit readout line as the main loss channel.

\begin{figure}[h]
    \centering
    \includegraphics[width=0.45\textwidth]{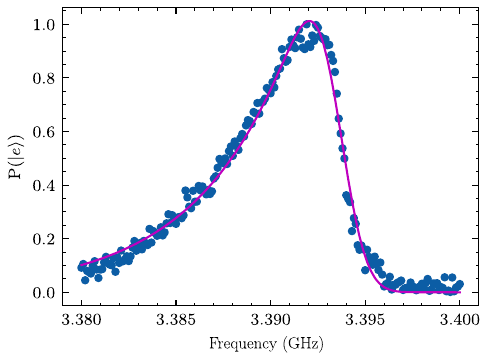}
\caption{Probability to find the qubit in its excited state $|e\rangle$ after playing a selective pi-pulse, as a function of the frequency of the pulse. Blue points: data, purple line: fit to a model including a dispersive pull by the memory resonator with $\chi/2\pi=1.5$~MHz and a memory temperature $T=600\pm50$~mK.}
    \label{fig:qubit_2tone_half_flux}
\end{figure}

We cannot proceed the same way on the flux line, which is not connected to a dedicated microwave measurement setup. Nonetheless, we have been able to measure the memory mode temperature in a different run where a reflective low-pass filter on the flux line had been removed, allowing the flux line to thermalize to the temperature of its last attenuator, anchored to a cold plate positioned between the still and mixing chamber stages. In that run, the qubit spectroscopy (Fig.~\ref{fig:qubit_2tone_half_flux}) revealed a wide asymmetric peak, indicating that the memory was in a thermal state with a large population dispersively pulling on the qubit~\cite{schuster2007resolving}. Fitting this peak yields an elevated temperature of 600 mK for the memory mode, indicating that its coupling to the flux line imposes its equilibrium temperature. This also indicates that stray coupling of the memory to the flux line is the main loss channel.

The fact that the internal losses probably do not contribute to the low memory quality factor is also confirmed by the very long lifetime of the transmon qubit of about 100 $\mu$s, which hints that our nanofabrication is good enough to achieve higher internal quality factors for resonators than the total observed quality factor.

The origin of this spurious coupling to the flux line can be attributed to the design of its termination, which was supposed to yield $Q_{flux}=10^6$ in the case where the flux line port is matched precisely to 50 ohms. Finite element simulations performed with Ansys HFFS show that even a modest impedance mismatch at the port can bring $Q_{tot}$ down to as low as $10^5$ due to a standing wave mode in the flux line.

The same simulations show that a better design of the flux line, including a shunting termination close to the SQUID, would suppress this loss channel, allowing the memory to reach much longer lifetimes.

\subsection{Measuring memory lifetime under two-photon dissipation}
\label{sec:memory_lifetime}
To measure the memory lifetime when the two-photon swap process is on, we measure the evolution of the Wigner function at the origin of phase space, i.e.\ $2/\pi$ times the memory photon-number parity, which is unchanged by two-photon loss. Fitting $W(0, \tau)$ with the expression expected from a decaying coherent state starting in $\alpha_0=\sqrt{n_0}$: $W(0,\tau)=2/\pi\times e^{-2n_0e^{-\kappa_a\tau}}$ directly yields the memory lifetime.

\subsection{Losses due to parasitic 1-photon conversion processes}
\label{sec:parasitic_1photon}
As discussed in the main text, the dc-voltage-biased SQUID can also swap a single memory photon with the electromagnetic modes of its environment at frequencies $\omega_a-\omega_J$ and $\omega_a+\omega_J$. We derive here the resulting memory loss rate $\kappa_a^\mathrm{extra}$ by treating the SQUID as a classical inductance modulated at the Josephson frequency and computing the small-signal admittance it presents to the memory.
\paragraph{Classical small-signal admittance of the biased junction.}
We consider the SQUID as a single junction of critical current $I_0$, in series with the impedance $Z(\omega)$ seen from its terminals, and biased by a dc voltage $V$ on top of which a small ac voltage $v(t)=\tilde v(\omega)e^{i\omega t}+\mathrm{c.c.}$ is applied. Writing $\varphi_0=\hbar/2e$ for the reduced flux quantum, the current $I(t)=I_0\sin\varphi(t)$ and the phase $\varphi$ obey $\varphi_0\dot\varphi=V_J$, where $V_J=V+v(t)-(\tilde Z\circledast I)(t)$ is the voltage across the junction and $\tilde Z$ the impulse response of $Z(\omega)$. The dc bias sets $\varphi(t)=\omega_J t+\varepsilon(t)$ with $\omega_J=2eV/\hbar$, and the ac part reads in the frequency domain
\begin{equation}
\tilde\varepsilon(\omega')=\frac{\tilde v(\omega')-Z(\omega')\tilde I(\omega')}{i\omega'\varphi_0}
=\frac{\tilde V_J(\omega')}{i\omega'\varphi_0}.
\label{eq:eps_phasor}
\end{equation}
In the small-signal limit $|\varepsilon|\ll1$, valid as long as all ac voltages remain small compared to $\varphi_0\omega'$, we expand
\begin{equation}
I(t)\simeq I_0\sin(\omega_J t)+I_0\cos(\omega_J t)\,\varepsilon(t).
\label{eq:current_expansion}
\end{equation}
The first term is the ac Josephson current at $\omega_J$. It does not feed back on the memory at this order and is responsible for the driving of low-frequency modes discussed in Sec.~\ref{sec:dephasing_bias_noise}. The second term is a mixing term. Since $\cos(\omega_J t)=(e^{i\omega_Jt}+e^{-i\omega_Jt})/2$, an ac phase at frequency $\omega$ generates currents at the idler frequencies $\omega_\pm=\omega\pm\omega_J$, and reciprocally a phase at $\omega_\pm$ generates a current at $\omega$. Applying harmonic balance to Eq.~\eqref{eq:current_expansion} and keeping only the frequencies $\omega$ and $\omega_\pm$, we obtain
\begin{align}
\tilde I(\omega_\pm)&=\frac{I_0}{2}\tilde\varepsilon(\omega),
\label{eq:hb_idler}\\
\tilde I(\omega)&=\frac{I_0}{2}\left[\tilde\varepsilon(\omega_-)+\tilde\varepsilon(\omega_+)\right].
\label{eq:hb_signal}
\end{align}
Terms at $\omega\pm2\omega_J$ are neglected because $Z$ is not resonant there. No voltage is applied at the idler frequencies, so that Eq.~\eqref{eq:eps_phasor} gives $\tilde\varepsilon(\omega_\pm)=-Z(\omega_\pm)\tilde I(\omega_\pm)/(i\omega_\pm\varphi_0)$. Inserting Eqs.~\eqref{eq:hb_idler} and~\eqref{eq:eps_phasor} into Eq.~\eqref{eq:hb_signal}, the current at $\omega$ becomes proportional to the voltage $\tilde V_J(\omega)$ across the junction. The junction therefore behaves for the signal at $\omega$ as a linear admittance
\begin{equation}
Y_J(\omega,\omega_J)=\frac{\tilde I(\omega)}{\tilde V_J(\omega)}
=\frac{1}{4L_J^2}\left[\frac{Z(\omega-\omega_J)}{\omega(\omega-\omega_J)}+\frac{Z(\omega+\omega_J)}{\omega(\omega+\omega_J)}\right],
\label{eq:YJ}
\end{equation}
with $L_J=\varphi_0/I_0=\varphi_0^2/E_J$ the Josephson inductance. The dissipative response of the junction at $\omega$ is thus entirely set by the environment impedance at the two idler frequencies. When $\omega_J<\omega$, both idler frequencies are positive and both terms of Eq.~\eqref{eq:YJ} have a positive real part for a passive environment, so that the junction acts as a positive resistance and induces losses. When $\omega_J>\omega$, the first idler frequency is negative and $Z(\omega-\omega_J)=Z^*(\omega_J-\omega)$ divided by a negative frequency yields a negative resistance, which is the gain mechanism of the inelastic-Cooper-pair-tunneling amplifier~\cite{Jebari2018,martel2025influence}. At the 2-to-1 working point of this work, $\omega_J/2\pi=0.53~\mathrm{GHz}$ is well below $\omega_a/2\pi=4.04~\mathrm{GHz}$ and we are in the purely dissipative regime.

\paragraph{From admittance to memory loss rate.}
The SQUID sits at a voltage antinode of the memory resonator (Fig.~\ref{fig:device}). A small admittance $Y$ shunting a quarter wavelength resonator of frequency $\omega_a$ at its voltage antinode changes its damping rate according to
\begin{equation}
\kappa_a^\mathrm{extra}=\frac 4  \pi\omega_a Z_a\,\mathrm{Re}\,Y(\omega_a),
\label{eq:shunt_admittance}
\end{equation}
where $Z_a$ is the effective characteristic impedance of the memory mode. Inserting Eq.~\eqref{eq:YJ} evaluated at $\omega=\omega_a$ into Eq.~\eqref{eq:shunt_admittance} and using $L_J=\hbar^2/(4e^2E_J)$, we find
\begin{equation}
\boxed{\;
\kappa_a^\mathrm{extra}
=\frac{4\phi_a^2}{R_Q}\left(\frac{E_J}{\hbar}\right)^2
\left[\frac{\mathrm{Re}\,Z(\omega_a-\omega_J)}{\omega_a-\omega_J}
+\frac{\mathrm{Re}\,Z(\omega_a+\omega_J)}{\omega_a+\omega_J}\right]
\;}
\label{eq:kappa_extra}
\end{equation}
The rate scales as $E_J(\Phi)^2\propto\cos^2(\pi\Phi/\Phi_0)$, as observed in Fig.~\ref{main-fig:fig4}b.

\begin{figure}[h]
    \centering
    \includegraphics[width=0.9\textwidth]{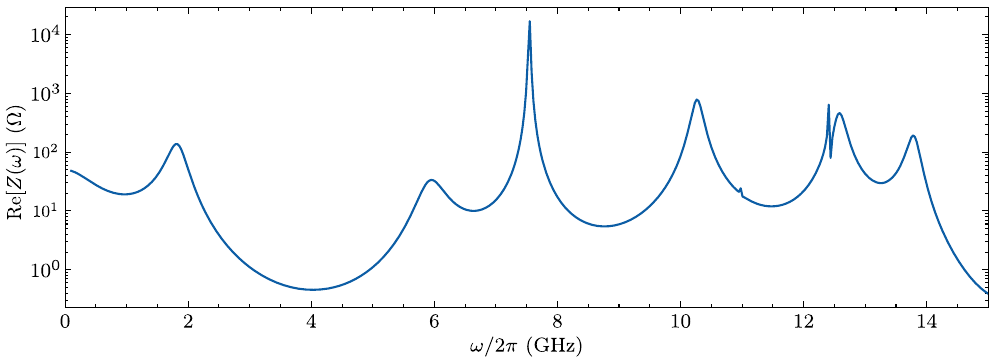}
\caption{As-designed real part of the impedance in parallel to the SQUID $\mathrm{Re}[Z(\omega)]$ computed from an Ansys HFSS simulation.}
    \label{fig:bragg_mirror_full}
\end{figure}

\paragraph{Comparison with the measured $E_J^2$ scaling.}
At the 2-to-1 working point, the idler frequencies are $(\omega_a-\omega_J)/2\pi=3.51~\mathrm{GHz}$ and $(\omega_a+\omega_J)/2\pi=4.57~\mathrm{GHz}$, both within the stop band of the stepped impedance filter. A linear fit of $\kappa_a$ against $\cos^2(\pi\Phi/\Phi_0)$ yields a slope {$2\pi\times2.9~\mathrm{MHz}$} per unit $\cos^2(\pi\Phi/\Phi_0)$, i.e. $\kappa_a^\mathrm{extra}/2\pi\simeq{2.9~\mathrm{MHz}}\times(E_J/E_J^\mathrm{max})^2$. Assuming a common value of $\mathrm{Re}\,Z$ at the two idler frequencies, Eq.~\eqref{eq:kappa_extra} with $\phi_a=0.18$ and $E_J^\mathrm{max}/h=13~\mathrm{GHz}$ translates this slope into $\mathrm{Re}\,Z(\omega_a\pm\omega_J)\simeq{1.8~\Omega}$. This value is a few times larger than the $0.5~\Omega$ predicted in this frequency range by the finite element simulation of the as-designed filter (Fig.~\ref{fig:bragg_mirror_full}). We attribute this discrepancy to nanofabrication uncertainties on the line widths and gaps of the filter, which degrade its rejection. The same measurements thus provide an in situ characterization of the residual impedance seen by the SQUID inside the stop band.

\begin{figure}[h]
    \centering
    \includegraphics[width=0.45\textwidth]{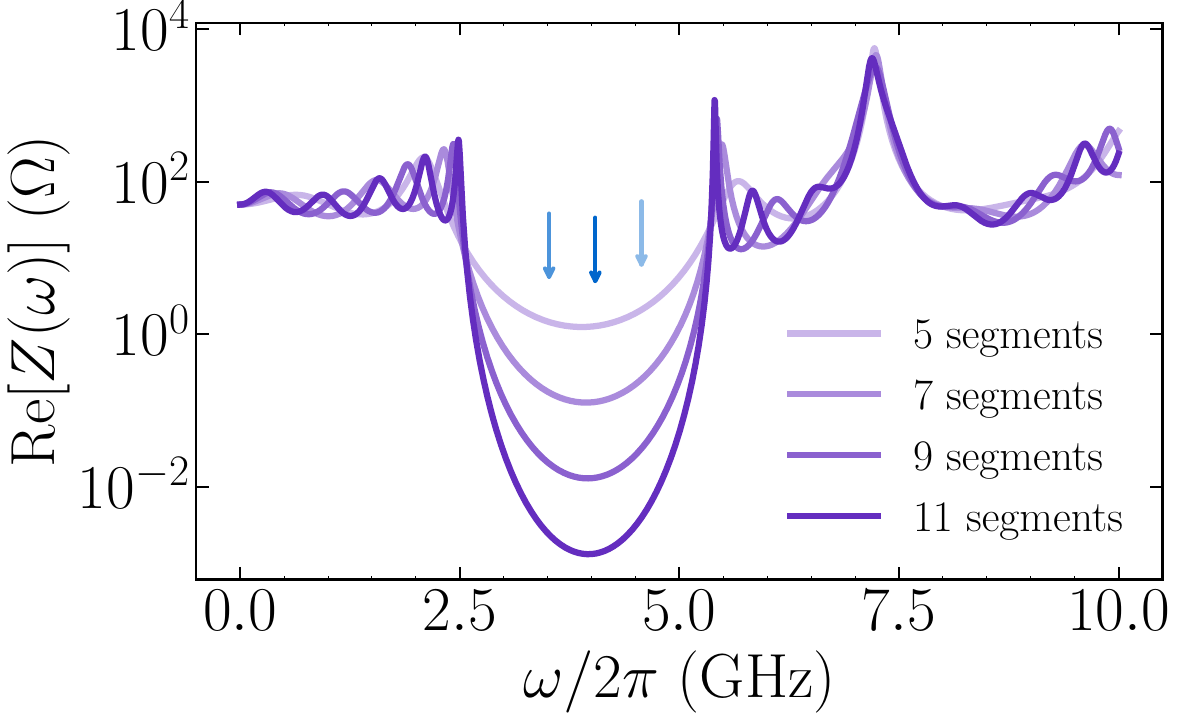}
\caption{$\mathrm{Re}[Z(\omega)]$ computed from a circuit model of the chip as-fabricated. The current device corresponds to the "5 segments" label. Increasing the number of segments in the stepped impedance filter reduces both the leakage of the memory to the bias line mode at its frequency $\omega_a$ (dark blue arrow), and the losses due to parametric swaps with modes at $\omega_a\pm \omega_J$ (light blue arrows).}
    \label{fig:bragg_mirror_ReZ_vs_freq}
\end{figure}

At large enough $E_J$, the memory lifetime will always be limited by these parasitic conversion processes, since both $\kappa_a^\mathrm{extra}$ and the engineered rate $\kappa_2=4g_2^2/\kappa_b$ scale as $E_J^2$ (in the adiabatic elimination regime). Their ratio,
\begin{equation}
\frac{\kappa_a^\mathrm{extra}}{\kappa_2}
=\frac{4\kappa_b}{R_Q\,\phi_a^2\phi_b^2}
\sum_\pm\frac{\mathrm{Re}\,Z(\omega_a\pm\omega_J)}{\omega_a\pm\omega_J},
\label{eq:ratio_extra_kappa2}
\end{equation}
is independent of $E_J$ and can only be reduced by lowering $\mathrm{Re}\,Z$ at the idler frequencies. This is the purpose of the stepped impedance filter, designed to reduce $\mathrm{Re}[Z(\omega)]$ in a $2~\mathrm{GHz}$ range around $\omega_a$ (Fig.~\ref{fig:bragg_mirror_full}). We checked with a circuit model of the device that increasing the number of segments of the filter would reduce $\mathrm{Re}[Z(\omega)]$ by several orders of magnitude at $\omega_a\pm\omega_J$ (Fig.~\ref{fig:bragg_mirror_ReZ_vs_freq}), suppressing these extra losses accordingly.

\subsection{Dephasing due to bias voltage noise}
\label{sec:dephasing_bias_noise}
In this section, we derive the fit function that was used in Fig.~4c of the main text. In the general case, biasing the SQUID at voltage $V=\hbar\omega_J/2e$ creates an oscillating current $I(t)=\frac{E_J}{\phi_0}\sin(\omega_J t)$, which can drive a mode $c$ of the circuit provided its frequency $\omega_c$ is close to $\omega_J$. Including explicitly this mode in the Hamiltonian of the circuit and performing a rotating wave approximation up to second order (as in appendix H in \cite{aissaoui_cat-qubit-stabilization_2026}), we find that the displacement of mode $c$ induces an AC stark shift of the memory mode frequency
\begin{equation}
\delta \omega_a(\omega_J) = \frac{E_J^2\phi_a^2\phi_c^2}{2\hbar^2}\frac{4\kappa_c}{4(\omega_c-\omega_J)^2+\kappa_c^2},
\end{equation}
with $\phi_c$ the zero point fluctuations of mode $c$ and $\kappa_c$ its inverse lifetime. Sweeping $\omega_J$ across $\omega_c$ in a range of the order of $\kappa_c$ thus induces a positive then negative shift of the memory frequency.
We also find that if $\omega_J \approx 2\omega_c$, the SQUID creates a 2-photon drive on mode $c$, resulting again in an AC stark shift of the memory with
\begin{equation}
\delta \omega_a(\omega_J) = \frac{E_J^2\phi_a^2\phi_c^4}{4\hbar^2}\frac{\kappa_c}{(2\omega_c-\omega_j)^2+\kappa_c^2}.
\end{equation}
These equations are perturbative in the $c$ mode population, and thus do not hold in all the range of $E_J$ values that we explore. Notably we find that for large $E_J$ they overestimate the magnitude of the memory shift. Nevertheless, the position and width of the shifts as a function of the bias $V$ are well reproduced by the model, even at larger $E_J$.
We fit the memory spectroscopy of Fig.~\ref{main-fig:fig4}c by summing the contributions of four processes:
\begin{itemize}
    \item the direct driving of a mode at $\omega/2\pi=227$ MHz, identified as a standing wave resonance in the bias line (Sec.~\ref{sec:sw_mode});
    \item a 2-photon drive on the same mode, occurring at $2eV/\hbar=2\times227$~MHz;
    \item the dispersive shift due to the 2-to-1 photon coupling to the buffer mode~\cite{lescanne2020exponential};
    \item the direct driving of an undetermined mode at $\omega/2\pi=766$~MHz.
\end{itemize}
Note that due to the finite population renormalization discussed earlier, we attribute an arbitrary amplitude to these four shifts.

Next, we numerically compute the derivative of $\delta \omega_a(V)$ from spectroscopies measured at different values of $E_J$ to obtain the sensitivity of the memory frequency to bias voltage. Multiplying this sensitivity by the value of the bias voltage noise $\sigma_V=3.1$~nV, we obtain an estimate of the memory dephasing rate which qualitatively agrees with the dephasing extracted from fitting the Wigner function dynamics in Sec.~\ref{sec:deflate_numerics} (see Fig.~\ref{main-fig:fig4}c inset).

\section{Numerical analysis of the two-photon dissipation dynamics}
\label{sec:deflate_numerics}

This section describes how the maximum two-photon coupling rate
$g_2^{\mathrm{max}}$, the memory dephasing rate
$\kappa_\phi(\Phi)$, and a bound on the memory self-Kerr rate $K$ are extracted from
the Wigner tomography records, an example of which is shown in Fig.~\ref{main-fig:fig3}b.

Two datasets are used. The deflate traces of Fig.~\ref{main-fig:fig4}a provide
the memory population $\bar n(\tau)$ as a function of time at each flux point
and constrain $g_2^{\mathrm{max}}$ and the photon-number axis calibration.
Time-resolved Wigner
tomographies at $\bar n(0)\simeq 4$ constrain $\kappa_\phi$. The two datasets
are complementary: dephasing reduces the coherence involved in the two-photon
conversion and modifies the deflate dynamics, but its effect on the population
alone cannot be separated from a change in $g_2^{\mathrm{max}}$ or in the
photon-number axis calibration.
The phase-space distribution distinguishes these effects. We therefore
alternate the two fits until they agree in a self-consistent manner, feeding only
$\kappa_\phi(\Phi)$ from the second fit into the first
(Sec.~\ref{sec:fixed_point}). A separate large-amplitude Wigner movie taken at
a single flux bounds $K$ (Sec.~\ref{sec:kerr_bound}). Table~\ref{tab:parameters}
summarizes the fitted and fixed parameters.

\subsection{Model}
\label{sec:model}

The memory $\hat a$ and the buffer $\hat b$ are described by
\begin{equation}
\dot\rho
=
-\frac{i}{\hbar}[\hat H,\rho]
+\kappa_a(\Phi)\,\mathcal D[\hat a]\rho
+\kappa_b\,\mathcal D[\hat b]\rho
+2\kappa_\phi(\Phi)\,\mathcal D[\hat a^\dagger\hat a]\rho ,
\label{eq:deflate_master_equation}
\end{equation}
\begin{equation}
\frac{\hat H}{\hbar}
=
-\Delta_a\,\hat a^\dagger\hat a
-\frac{K}{2}\,\hat a^{\dagger2}\hat a^2
+ g_2(\Phi)\left(\hat a^2\hat b^\dagger+\hat a^{\dagger2}\hat b\right).
\label{eq:wigner_hamiltonian}
\end{equation}

The memory is initialized in a coherent state and the buffer in vacuum. The
Hamiltonian is given by Eq.~\eqref{main-Hamiltonian} for $n=2$, written in the
rotating frame, with additional terms for the memory self-Kerr and detuning.
Using the flux dependence of the Josephson energy established in
Sec.~\ref{sec:squid_asymmetry}, the coupling rate is
\begin{equation}
g_2(\Phi)
=
g_2^{\mathrm{max}}\sqrt{\cos^2(\pi\Phi/\Phi_0)+\epsilon^2\sin^2(\pi\Phi/\Phi_0)} ,
\label{eq:deflate_model}
\end{equation}
where the SQUID asymmetry $\epsilon=0.016$ sets the residual coupling near
half flux. The memory thermal population is set to zero;  we checked that its measured value
$n_{\mathrm{th}}^a=0.005$ had no visible impact on the value of the parameters we extract using these simulations.

The self-Kerr rate is fixed to the upper bound determined at zero $E_J$,
\begin{equation}
\frac{K}{2\pi}=3~\mathrm{kHz}.
\label{eq:kerr_fixed}
\end{equation}

Kerr effect, detuning, and dephasing preserve the diagonal of $\rho$, so they do not
directly change the population. Their effect on $\bar n(\tau)$ enters through
the $\hat a^2$ coherence involved in the two-photon conversion. The Kerr rate is
held at the value of Eq.~\eqref{eq:kerr_fixed}, its effect on the deflation
dynamics being negligible. The detuning is set to zero here; it is not
independently determined at these flux points
(Sec.~\ref{sec:wigner_dephasing}). Dephasing suppresses the same coherence
significantly and is included in the deflate model, using the values extracted
from the Wigner tomographies (Sec.~\ref{sec:fixed_point}).

\begin{table}[htb!]
\centering
\begin{tabular}{lll}
\hline\hline
Parameter & Value & Determination \\
\hline
\multicolumn{3}{l}{\emph{Fitted}} \\
$g_2^{\mathrm{max}}/2\pi$
    & $18.2\pm0.5~\mathrm{MHz}$
    & Deflate fit, Sec.~\ref{sec:global_fit} \\
$A$
    & $1.131\pm0.009$
    & Deflate fit, Sec.~\ref{sec:global_fit} \\
$\kappa_\phi(\Phi)$
    & Table~\ref{tab:flux_points}
    & Wigner fit, Sec.~\ref{sec:wigner_dephasing} \\
\hline
\multicolumn{3}{l}{\emph{Fixed}} \\
$\kappa_a(\Phi)$
    & Table~\ref{tab:flux_points}
    & Independent extraction, Sec.~\ref{sec:memory_lifetime} \\
$\kappa_b/2\pi$
    & $85\pm5~\mathrm{MHz}$
    & Buffer impedance, Sec.~\ref{sec:buffer_impedance} \\
$\epsilon$
    & $0.016$
    & SQUID asymmetry, Sec.~\ref{sec:squid_asymmetry} \\
$K/2\pi$
    & $3~\mathrm{kHz}$
    & Upper bound of main text; Sec.~\ref{sec:kerr_bound} \\
$\Delta_a$
    & $0$
    & Deflate fit; fitted in Sec.~\ref{sec:wigner_dephasing} \\
\hline\hline
\end{tabular}
\caption{Parameters used in the two-mode model. Fitted parameters are obtained
from the deflate and Wigner analyses described in
Secs.~\ref{sec:global_fit} and~\ref{sec:wigner_dephasing}. Fixed parameters
are taken from independent measurements or set as stated in the text.}
\label{tab:parameters}
\end{table}

\subsection{Fit of the deflate traces}
\label{sec:global_fit}

\paragraph{Data.}
The dataset contains 29 traces of $\bar n(\tau)$, taken at the ten flux points
of Table~\ref{tab:flux_points}, spanning
$0.016\leq\left|\cos(\pi\Phi/\Phi_0)\right|\leq0.439$ and initial populations
$0.75\leq\bar n(0)\leq7.52$.

\paragraph{Memory loss.}
The memory loss is not a free parameter of this fit. At all ten flux points
$\kappa_a(\Phi)$ is taken from the independent extraction of
Sec.~\ref{sec:memory_lifetime}, linearly interpolated onto the fitted flux
points and kept fixed (Table~\ref{tab:flux_points}).

\paragraph{Photon-number axis.}
As explained in Sec.~\ref{sec:measuring_memory}, the measured Wigner functions
must be rescaled by a flux-dependent factor. The vacuum tomographies provide
the shape of this calibration, but do not reproduce the observed
$\bar n(\infty)$. We therefore keep the measured flux dependence and fit a
single global scale factor to the data.

\begin{equation}
c(\Phi)=A\,\hat c_{\mathrm{vac}}(\Phi),
\label{eq:deflate_cphi}
\end{equation}

where $\hat c_{\mathrm{vac}}(\Phi)$ is the vacuum calibration interpolated to
the fitted flux points. Eq.~\eqref{eq:wigner_integral} gives
\[
\int|\beta|^2W(\beta)\,\mathrm{d}^2\beta=\bar n+\tfrac12,
\]
so rescaling the phase-space axes by $\sqrt{c}$ rescales this quantity by $c$,
giving
\begin{equation}
n_{\mathrm{corr}} =
c(\Phi)\left(n_{\mathrm{raw}}+\tfrac12\right)-\tfrac12,
\qquad
\beta_{\mathrm{corr}} = \sqrt{c(\Phi)}\,\beta_{\mathrm{raw}} .
\label{eq:deflate_axis_correction}
\end{equation}

The scale factor $A$ and the coupling $g_2^{\mathrm{max}}$ are strongly
correlated because, in the mean-field limit,
\(\dot n\propto -(g_2^{\mathrm{max}})^2 n^2\), so a rescaling of the photon
number can be compensated by a change in the coupling rate. The independently
measured single-photon loss helps distinguish the two through the late-time
dynamics. At low photon numbers, the exact two-photon dynamics also deviate
from the mean-field $n^2$ scaling and provide an additional constraint. The fit
gives
\begin{equation}
A=1.131\pm0.009.
\label{eq:deflate_cphi_fit}
\end{equation}

\begin{figure}[htb!]
\centering
\includegraphics[width=0.5\linewidth]{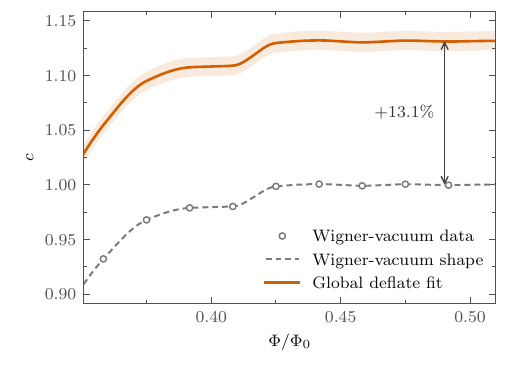}
\caption{Photon-number calibration factor $c$ as a function of flux. Circles
show the vacuum-tomography calibration and the dashed line its interpolant
($A=1$). The orange line and band show the global fit, giving
$A=1.131\pm0.009$ ($1\sigma$).}
\label{fig:deflate_axis_calibration}
\end{figure}

This is a $13.1\%$ offset from the vacuum calibration
(Fig.~\ref{fig:deflate_axis_calibration}). There is no independent measurement
of the photon-number axis at the operating point, so the fitted offset may
contain both calibration and model contributions. We take the full range
$A\in[1,1.131]$ as a one-sided systematic uncertainty on the fitted parameters.
Fixing $A=1$ shifts $g_2^{\mathrm{max}}/2\pi$ upward by $2.6~\mathrm{MHz}$ and
raises $\chi^2_\nu$ from $1.99$ to $2.44$.

\paragraph{Fitting procedure.}
The traces are fitted by weighted least squares, using one noise estimate per
trace. We estimate the noise from local pseudo-residuals and take their median
as the standard deviation~\cite{gasser1986residual}. The residuals are computed
in the raw measurement space: instead of correcting the data with
Eq.~\eqref{eq:deflate_axis_correction}, we apply its inverse to the simulated
trace, so that the fitted scale factor $A$ does not change the weighting of the
measurement noise.

The master equation is integrated with
\texttt{dynamiqs}~\cite{guilmin2025dynamiqs}, and the parameters are optimized with \texttt{scipy} using \texttt{optimize.least\_squares}, taking the Jacobians computed with \texttt{JAX}~\cite{jax2018github}. Statistical uncertainties are obtained from the Gauss--Newton covariance matrix.

\begin{figure*}[htb!]
\centering
\includegraphics[width=\textwidth]{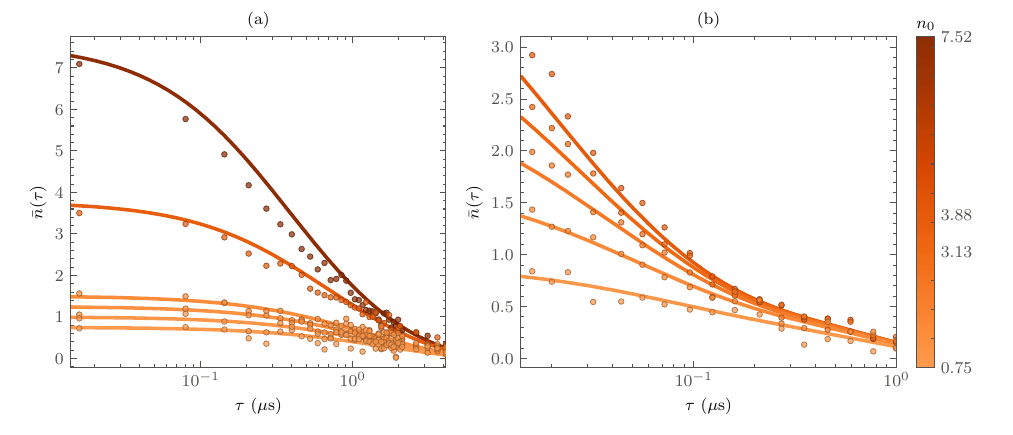}
\caption{Mean memory occupation versus time at
$\left|\cos(\pi\Phi/\Phi_0)\right|=0.036$ (\textbf{a}) and $0.243$
(\textbf{b}), the two flux points at which the initial photon number was swept.
Circles: data. Lines: simultaneous fit to all 29 traces
(Sec.~\ref{sec:global_fit}) at $\kappa_b/2\pi=85~\mathrm{MHz}$. The color
encodes the initial photon number $\bar n(0)$.}
\label{fig:deflate_g2_fit}
\end{figure*}

\paragraph{Result.}
With $\kappa_\phi(\Phi)$ fixed to the values of
Sec.~\ref{sec:wigner_dephasing}, the fit yields
\begin{equation}
\frac{g_2^{\mathrm{max}}}{2\pi}
=18.2\pm0.5\,~\mathrm{MHz},
\label{eq:deflate_g2_result}
\end{equation}
with $\chi^2_\nu=2.0$ and an rms residual of $0.16$ photons
(Fig.~\ref{fig:deflate_g2_fit}). The $\pm0.5~\mathrm{MHz}$ uncertainty combines
the statistical uncertainty of $0.16~\mathrm{MHz}$ with the uncertainty from
the buffer linewidth: varying $\kappa_b$ by $\pm5~\mathrm{MHz}$ changes
$g_2^{\mathrm{max}}/2\pi$ by $0.46~\mathrm{MHz}$. 

This value is consistent with the circuit parameters of
Table~\ref{tab:device_parameters}. Using
$g_2=E_J\phi_a^2\phi_b/4\hbar$ with $E_J^{\mathrm{max}}/h=13~\mathrm{GHz}$,
$\phi_a=0.18$, and $\phi_b=0.21$ gives
$g_2^{\mathrm{max}}/2\pi=22.1~\mathrm{MHz}$, about $21\%$ above the fitted
value, an acceptable discrepancy given that the zero-point phases are design values rather than measurements.
Since the data stop at $\left|\cos(\pi\Phi/\Phi_0)\right|=0.439$,
Eq.~\eqref{eq:deflate_g2_result} is an extrapolation of
Eq.~\eqref{eq:deflate_model} to zero flux. The directly constrained rate at
the strongest measured coupling point is $g_2/2\pi=8.0~\mathrm{MHz}$.

\subsection{Fit of the Wigner tomographies}
\label{sec:wigner_dephasing}

\paragraph{Data.}
The tomographies are taken at $\bar n(0) = 4.1$ and provide 20 frames per
flux point, between $16~\mathrm{ns}$ and $1.6~\mu\mathrm{s}$, on a
$51\times51$ grid in phase space. The frames are background-corrected and
normalized to unit integral on the corrected grid
$\beta_{\mathrm{corr}}=\sqrt c\,\beta_{\mathrm{raw}}$.

\paragraph{Fitting procedure.}
We trace out the buffer, calculate the memory Wigner function, and minimize the
unweighted sum of squared residuals over all pixels and frames. We use an
unweighted fit because the tomography uses a fixed number of shots per point,
and the resulting pixel variance varies only weakly across the measured grid.
At each flux point, the free parameters are $\kappa_\phi\geq0$ and the memory
detuning $\Delta_a$ of Eq.~\eqref{eq:wigner_hamiltonian}; the remaining model
parameters are fixed. The Wigner distribution distinguishes the two:
dephasing suppresses phase coherence, whereas detuning produces a coherent
rotation of the distribution.

\begin{figure*}[htb!]
\centering
\includegraphics[width=\textwidth]{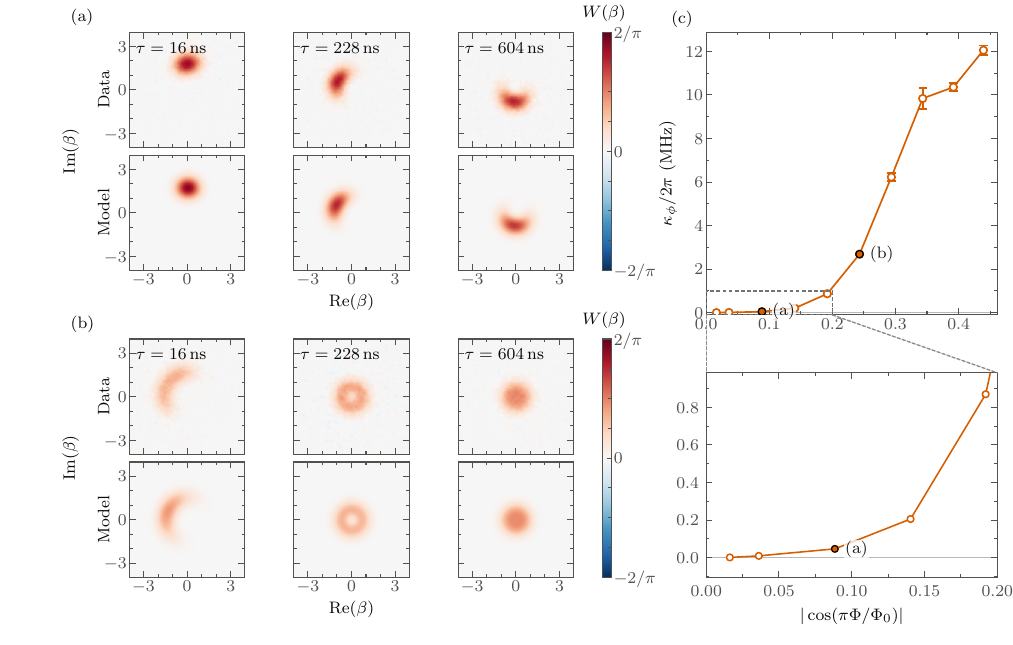}
\caption{\textbf{a,b,} Measured (top) and fitted (bottom) Wigner functions at
$\left|\cos(\pi\Phi/\Phi_0)\right|=0.088$ and $0.243$, at the times indicated.
At the weaker coupling the distribution stays compact as it decays, while at
the stronger one it is azimuthally scrambled into a ring by $\tau=228~\mathrm{ns}$.
\textbf{c,} Fitted $\kappa_\phi/2\pi$ against
$\left|\cos(\pi\Phi/\Phi_0)\right|$, with error bars showing the total
uncertainty of Sec.~\ref{sec:uncertainties}. The lower panel is a magnification
of the weak-coupling region. Filled markers indicate the two flux points shown
in \textbf{a} and \textbf{b}.}
\label{fig:wigner_dephasing}
\end{figure*}

\paragraph{Result.}
The dephasing rate increases by more than four orders of magnitude across the
flux range, from
\(\kappa_\phi/2\pi=8.8\times10^{-4}~\mathrm{MHz}\) at
\(\left|\cos(\pi\Phi/\Phi_0)\right|=0.016\) to
\(12.1\pm0.2~\mathrm{MHz}\) at \(0.439\)
(Fig.~\ref{fig:wigner_dephasing}, Table~\ref{tab:flux_points}). These values
are shown as squares in the inset of Fig.~\ref{main-fig:fig4}c and compared
with the rate
\(\kappa_\phi=(\partial\delta\omega/\partial\omega_J)\,\sigma_J\) expected
from the measured memory pull and residual bias-voltage noise. The quoted
uncertainties are dominated by the calibration uncertainties of
Sec.~\ref{sec:uncertainties}. At the weakest coupling, the coherence decay over
the movie duration is small, so that value is better read as an upper bound.
The fitted $\Delta_a$ stays below a few MHz except at the strongest
couplings, where it grows and is poorly determined.

\begin{table}[htb!]
\centering
\begin{tabular}{ccc}
\hline\hline
$\left|\cos(\pi\Phi/\Phi_0)\right|$ & $\kappa_a/2\pi$ (MHz) & $\kappa_\phi/2\pi$ (MHz) \\
\hline
0.439 & $0.815\pm0.045$ & $12.1\pm0.2$ \\
0.392 & $0.598\pm0.050$ & $10.4\pm0.2$ \\
0.343 & $0.476\pm0.036$ & $9.8\pm0.5$ \\
0.294 & $0.340\pm0.021$ & $6.2\pm0.2$ \\
0.243 & $0.221\pm0.0066$ & $2.69\pm0.06$ \\
0.192 & $0.144\pm0.0031$ & $0.870\pm0.009$ \\
0.140 & $0.094\pm0.0016$ & $0.205\pm0.005$ \\
0.088 & $0.084\pm0.0011$ & $(4.64\pm0.19)\times10^{-2}$ \\
0.036 & $0.074\pm0.0006$ & $(9.2\pm0.5)\times10^{-3}$ \\
0.016 & $0.064\pm0.0003$ & $(8.8\pm3.3)\times10^{-4}$ \\
\hline\hline
\end{tabular}
\caption{Memory loss $\kappa_a$ and dephasing $\kappa_{\phi}$ at each flux point. All $\kappa_a$ values
come from the independent extraction of Sec.~\ref{sec:memory_lifetime} and
are held fixed in the fits. For the points with $ \Phi/\Phi_0 =$ $0.088$ and $0.036$, where the fit is not available, we linearly interpolated between the other fitted points. The uncertainties on $\kappa_\phi$ include the contributions
described in Sec.~\ref{sec:uncertainties}.}
\label{tab:flux_points}
\end{table}

\subsection{Self-consistency}
\label{sec:fixed_point}

The deflate fit is initialized with $\kappa_\phi=0$. The resulting parameters
are then used for the Wigner fits, whose extracted $\kappa_\phi(\Phi)$ is fed
back into the deflate fit. This procedure is repeated until the fitted
parameters change by less than their statistical uncertainties. Only
$\kappa_\phi(\Phi)$ is passed between the two fits.

At the strongest couplings, dephasing is fast enough to affect the deflation
dynamics significantly. Starting from a fit with $\kappa_\phi=0$, the first
iteration increases $g_2^{\mathrm{max}}/2\pi$ from $17.65$ to
$18.29~\mathrm{MHz}$, while $\chi^2_\nu$ decreases from $3.15$ to $1.96$.
A second iteration moves $g_2^{\mathrm{max}}/2\pi$ by $0.05~\mathrm{MHz}$,
well inside its statistical uncertainty, and leaves all other fitted parameters
unchanged at the quoted precision. Including dephasing therefore shifts the
extracted $g_2^{\mathrm{max}}/2\pi$ by $0.59~\mathrm{MHz}$.

\subsection{Uncertainties}
\label{sec:uncertainties}

The uncertainties quoted above combine the Gauss--Newton covariance of the fit
with the parameter shifts obtained by displacing each fixed input by
$\pm1\sigma$ and repeating the complete self-consistency procedure. The inputs
in Table~\ref{tab:parameters} without a stated uncertainty are not included in
this propagation. The photon-number-axis range $A\in[1,1.131]$ is treated
separately as a one-sided systematic uncertainty rather than included in
quadrature. The quoted $\kappa_\phi$ values use the fitted photon-number
calibration.

The mean residual drifts systematically with flux, with a slope of
$0.154\pm0.048$ photons per unit $\left|\cos(\pi\Phi/\Phi_0)\right|$, and this
drift is unchanged across the axis-calibration models considered. It indicates
a limitation of the two-mode model at the strongest couplings that is not
included in the quoted uncertainties.

\subsection{Bound on the memory self-Kerr}
\label{sec:kerr_bound}

A self-Kerr offsets the Fock state \(|n\rangle\) by a phase
\((K/2)n(n-1)\tau\) in a duration $\tau$, which shears the phase-space distribution at a rate that
grows with photon number. This deformation differs from the rigid rotation
caused by detuning and the phase-space broadening caused by dephasing, allowing
$K$ to be bounded even in the presence of strong dephasing. We use a
large-amplitude Wigner movie taken at
$\left|\cos(\pi\Phi/\Phi_0)\right|=0.036$, with 38 frames up to
$\bar n(0)=9.40$. The large displacement enhances the Kerr-induced deformation.

\begin{figure*}[htb!]
\centering
\includegraphics[width=\textwidth]{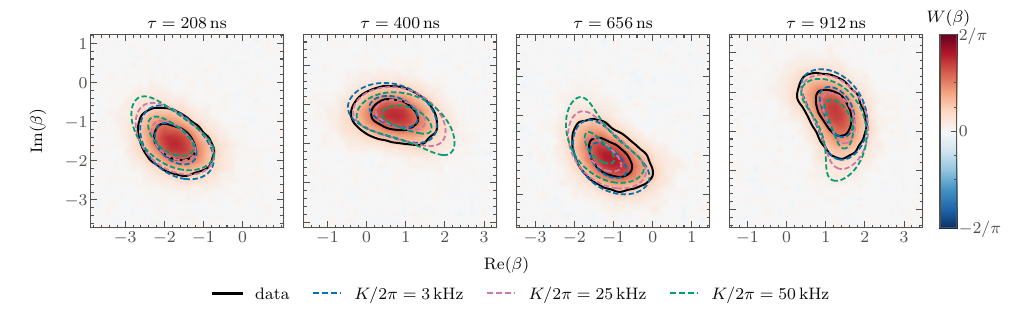}
\caption{Bound on the memory self-Kerr rate. Measured Wigner functions at
$\left|\cos(\pi\Phi/\Phi_0)\right|=0.036$ and $\bar n(0)\simeq9.4$, overlaid
with isolines of the data (solid) and of simulations at
$K/2\pi=3$, $25$, and $50~\mathrm{kHz}$ (dashed), drawn at $30\%$ and $70\%$
of the measured peak. The detuning and preparation phase are re-optimized at
each $K$. Larger $K$ elongates the distribution tangentially as the movie
proceeds.}
\label{fig:kerr_contours}
\end{figure*}

We evaluate the model for different values of $K$, keeping the remaining
physical parameters fixed. At each $K$, the detuning and preparation phase
$\varphi_0$ are re-optimized, since a detuning produces a rigid rotation that
could otherwise be confused with the Kerr shear. The agreement with the data
remains within the observed scatter up to $K/2\pi=10~\mathrm{kHz}$. It degrades
above that: at $25~\mathrm{kHz}$ the mismatch is clearly larger than the
scatter, and at $50~\mathrm{kHz}$ the simulated distribution develops the
crescent-shaped deformation visible in Fig.~\ref{fig:kerr_contours}. We
therefore obtain
\begin{equation}
\frac{K}{2\pi} \lesssim 10~\mathrm{kHz}
\quad\text{at }\left|\cos(\pi\Phi/\Phi_0)\right|=0.036,
\label{eq:kerr_bound}
\end{equation}
consistent with the value $K/2\pi=3~\mathrm{kHz}$ used in the model.

\section{Fabrication of the device}

We start from a 2 inch C-plane (0001) HEM-grown sapphire wafer, 430 µm thick, double-side-polished, supplied by UniversityWafer, Inc. The wafer is cleaned using successive baths of toluene, acetone, methanol, and isopropanol for one minute each, followed by another cleaning in a piranha solution (1 part hydrogen peroxide and 4 parts sulfuric acid) for 5 minutes. Finally, the wafer is annealed for 30 minutes at 1200°C under a nitrogen flow.
For the superconducting metal layer, 200~nm of tantalum is sputtered onto the wafer by STAR Cryoelectronics.

The large features of the device (control lines and resonators) are realized using e-beam lithography and reactive ion etching. These steps are performed at CEA Saclay. We use CF$_4$ gas at 20 sccm and
0.02 mbar to reactive ion etch the tantalum layer on the patterned
wafer. After stripping the resist, the sample is then cleaned in an NPM bath for 1 hour followed by
acetone and IPA baths. Finally, it undergoes an oxygen ashing step (3 minutes, 0.035 mbar, 200 W) to strip the resist residues. The wafer is then diced into individual chips.

We then use e-beam lithography on a single chip to fabricate the aluminum Josephson junctions on top of the tantalum circuit. These steps are performed at ENS Lyon. After cleaning in N-Methyl-2-pyrrolidone, acetone and isopropanol baths for 5 minutes each, a coating of hexamethyldisilazane is deposited using a vapor primer to promote resist adhesion. We then spincoat a PMGI-PMMA bilayer, baking each layer at 200°C for 10 minutes. We deposit a 10 nm layer of aluminum on the resist as a discharge layer for scanning electron microscope (SEM) lithography using a Plassys e-beam evaporator. We use a Zeiss SUPRA 55-VP microscope with 28 kV acceleration voltage to expose the PMMA layer, defining Dolan bridges. After removal of the aluminum discharge layer in a potassium hydroxide solution, the PMMA is developed in an MIBK-isopropanol 1-3 solution for 1 minute. The PMGI is then developed in a cold MF319 solution for 30 seconds, defining the bridge undercuts.
The chip is loaded into a Plassys evaporator and pumped overnight or until the loadlock pressure reaches $10^{-7}$~mbar. After argon ion-milling the tantalum surface to remove oxides, a first layer of 15 nm of aluminum is deposited at a 25° angle. The aluminum is left to oxidize in a 10 mbar oxygen atmosphere for 30 minutes, after which the loadlock is pumped again and a 100 nm aluminum layer is deposited at a -25 degrees angle, creating the Josephson junction. The resist mask is then lifted off in a 60°C NMP bath for 1 hour.

\bibliography{biblio}